\documentclass[12pt]{iopart}
\pdfoutput=1
\usepackage[english]{babel}
\usepackage{amssymb,amsfonts,latexsym,bm}
\usepackage[pdftex]{graphicx}
\usepackage{color}

\usepackage[colorlinks,bookmarks=false,citecolor=blue,linkcolor=red,urlcolor=blue]{hyperref}

\newcommand\be{\begin{equation}}
\newcommand\bea{\begin{eqnarray}}
\newcommand\eea{\end{eqnarray}}
\newcommand\ee{\end{equation}}

\begin{document}

\title{Non-equilibrium transport in $d$-dimensional non-interacting Fermi gases}
\author{Mario Collura, Gabriele Martelloni}
\address{Dipartimento di Fisica dell'Universit\`a di Pisa and INFN, 56127 Pisa, Italy}
\ead{collura@df.unipi.it, martelloni@df.unipi.it}
\date{\today}

\begin{abstract}
We consider a non-interacting Fermi gas in $d$ dimensions, both in the 
non-relativistic and relativistic case.
The system of size $L^{d}$ is initially prepared into two halves 
$\mathcal{L}$ and 
$\mathcal{R}$, each of them thermalized at two different 
temperatures, $T_{\mathcal{L}}$ and $T_{\mathcal{R}}$ respectively. 
At time $t=0$ the two halves are put in contact and the entire system is left to evolve unitarily.
We show that, in the thermodynamic limit, the time evolution of the particle and energy densities 
is perfectly described by a semiclassical approach which permits to analytically evaluate 
the correspondent stationary currents. 
In particular, in the case of non-relativistic fermions, 
we find a low-temperature behavior for the particle and energy 
currents which is independent from the dimensionality $d$ of the system,
being proportional to the difference $T_{\mathcal{L}}^{2}-T_{\mathcal{R}}^{2}$. 
Only in one spatial dimension ($d=1$), the results for the non-relativistic case
agree with the massless relativistic ones.
\end{abstract}

\section{Introduction}
In the last two decades, the improvement of the experimental techniques
has made it possible the experimental realization of the non-equilibrium dynamics 
of trapped ultra-cold atomic gases with high precision\cite{uc, kww06, tc07, tetal11, cetal12, getal11,shr12,rsb13}. One of the effects of such an experimental enhancement was to promote 
the theoretical studies of the non-equilibrium properties of many-body quantum systems.
In particular, the theoretical attention has been mainly focused both on the 
(generalized)thermalization mechanisms in  one-dimensional quantum systems
\cite{rdyo07,mussardo10-13,cc06,c06,cdeo08,bs08,
ce13,fe13,fcec13}  
(see Ref. \cite{revq} for a general review),
and on the out-of-equilibrium transport properties
\cite{araki,ogata,aschbacher,pk07,dvbd13,mm11,arrs98,ruelle98,jp02,
kp09,prosen11,kps13,cacdh13,CICC,KIM,SM}.

In particular, with respect to the transport properties in a non-equilibrium stationary-state (NESS),
 a very interesting situation is obtained by considering a gas of atoms
 initially splitted into two different packets, each of them prepared at given initial temperature. 
The gas is then released  and left to evolve freely. 

In this regard,  recent works in one spatial dimension have established, 
using Conformal Field Theory (CFT), an universal transport regime whenever 
two initially isolated critical systems are instantaneously brought  in contact \cite{bd12}.
This particular behavior has been inspected in the spin-$1/2$ XXZ quantum chain \cite{KIM}
by means of time-dependent Density Matrix Renormalization Group (tDMRG) algorithm
at finite temperature \cite{kbm12,kbm13,hkm13}. Interestingly, in Ref. \cite{KIM} it has been
found that the asymptotic energy current seems to be in agreement, within the numerical errors,
with the functional form $f(T_{\mathcal{L}})-f(T_{\mathcal{R}})$.
Nonetheless, recently, using a generalization of the Thermodynamic Bethe Anstaz (TBA) \cite{Zamo90}, 
the NESS energy transport for any integrable model of relativistic quantum field 
theory (IQFT) has been evaluated \cite{cacdh13}, and it has been found that the ``additivity'' property 
of the energy current, which holds in CFT, does not hold in general IQFT.  In this regard,
it has been suggested that the violation of the additivity property may not be visible in Ref. \cite{KIM}
since lower than the numerical precision.

These studies make it evident that the debate on the non-equilibrium 
properties in interacting systems is extremely promising and active. 

Inspired by this scenario and by the new results coming from the gauge gravity duality \cite{Hartnoll,MG,HolograficDoyon}, we extended the analysis carried out for $d=1$ in Ref. \cite{ck14} 
to a $d$-dimensional space, and for more general initial conditions.
\begin{figure}[t!]
\center\includegraphics[width=0.75\textwidth]{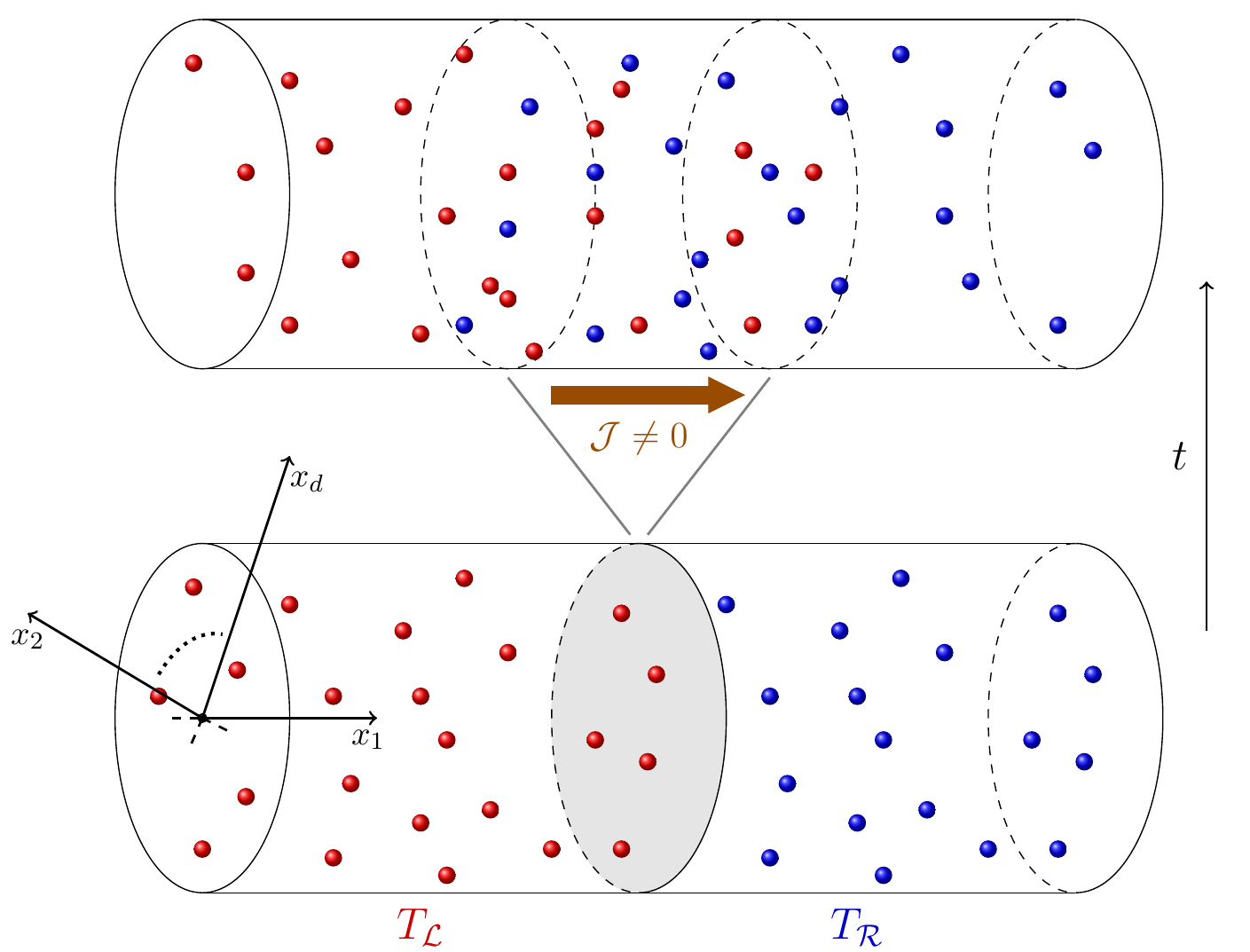}
\caption{Two Fermi gases prepared at different temperatures $T_\mathcal{L}$
and $T_{\mathcal{R}}$ are instantaneously joined together and left 
to evolve with a non-interacting Hamiltonian. In the thermodynamic limit and for large times
a current-carrying state opens across the interface.
The two different colored particles are just a graphical expedient in order to 
emphasize the two different temperatures.} 
\label{fig_sketch}
\end{figure}

Accordingly, in this paper we consider a local quantum quench, for a non-interacting Fermi gas,
along the spatial direction $x_{1}$, while in the other $d-1$ directions the geometry 
is left unchanged. After preparing the system in a tensor thermal state, i.e. a tensor product 
of two different thermal density matrices at two different temperatures 
($T_{\mathcal{L/R}}\equiv1/\beta_{\mathcal{L/R}}$) and chemical potentials ($\mu_{\mathcal{L/R}}$),
we leave it to evolve with a non-interacting Hamiltonian.
 In particular, for an infinite system, by means of a semiclassical description 
 \cite{akr08,cark12,wck13}, we fully characterize the time-evolution of the profiles 
 of the density of particles and of the energy density. From these profiles, we extract
 the analytic expression for the non-equilibrium currents.
 
We find that, for small temperatures ($T_{\mathcal{L/R}}\ll 1$), 
 the energy and the particle
  currents show an universal behavior on temperature 
$\sim (T_{\mathcal{L}}^{2}-T_{\mathcal{R}}^{2})$, independent on the dimensionality $d$. 
This result could be surprising and suggest a dimensional scaling violation.
Nevertheless, the correct dimensional scaling of the currents is restored 
thanks to the appropriate powers of the other scaling dimensions entering the problem, 
namely the mass of the particles and the chemical potential. 
 
Moreover,  we point out that only in one spatial dimension and at zero chemical potential 
the result agrees with the CFT predictions, due to the fact that, just for $d=1$,
  the mass disappears from the calculation. Na\"ively, one might think that, expanding the
  non-relativistic Fermi distribution for small temperatures, and keeping the linear behavior of the 
  dispersion relation around the Fermi surface, this should lead to the same relativistic result.
  Nevertheless, starting from a dispersion relation of non-relativistic particles, we argue that 
  expanding the observables at small temperatures, the CFT predictions cannot be recovered.
   
 The paper is organized as follows. In Sec. \ref{model_quench} we present the model and the
 post-quench Hamiltonian which governs the unitary evolution. In Sec. \ref{Obs} 
 we define the observables, i.e. the linear density of particle and the linear 
 density of energy along the quenched direction, as well as the particle current and the energy current.
  In Sec. \ref{PQD} we study the dynamics of the problem in the semiclassical approximation, and
  we show that the problem can be mapped to an analogous 1D problem; 
  we support the analytic results comparing them with exact numerical calculations. 
 In Sec. \ref{sec_currents} and \ref{RC} we finally give the analytic expression for the
 particle and energy currents in generic dimensions in the case, 
 respectively, of non-relativistic and relativistic dispersion relation. 
 Finally in Sec. \ref{Concl} we draw our conclusions.

 \section{Model and Quench}\label{model_quench}
We consider a non-relativistic quantum-field theory describing non-interacting spinless fermions
with mass $m$, in $d$ spatial dimensions. Since we are interested in the non-equilibrium stationary 
properties after a quantum quench along the direction $ x_1$, we separately consider the 
$x_1$ coordinate which is defined in $[-L/2,L/2]$. We introduced the system length 
$L$ along $ x_1$ in order to regularize infrared divergencies.
For the same reason, all the other coordinates $x_{i}$, with $i=2,3,\dots d$, 
are restricted in the symmetric domain $\mathcal{V}_{d-1}\equiv [-L/2,L/2]^{\otimes (d-1)}$. 
For brevity, whenever it will be not ambiguous,
we will consider the entire definition domain $\mathcal{V}\equiv [-L/2,L/2]\otimes \mathcal{V}_{d-1}$.
At the end, we will be interested in the thermodynamic limit (TD limit) $L\to\infty$ after having
integrated out all the others directions. 
In this way, the results will be valid even for a finite domain $\mathcal{V}_{d-1}$.

The post-quench Hamiltonian governing the unitary evolution of the system is a direct sum
of $d$ free hamiltonians, namely
\be\label{H}
\hat H = \int_{\mathcal{V}} \prod_{i=1}^{d} dx_i \, \hat\Psi^{\dag}({\bm x}) \left\{ -  \frac{1}{2m}\sum_{i=1}^{d}\partial^{2}_{x_i} \right\} \hat\Psi({\bm x}),
\ee
where the fields $\hat\Psi({\bm x})$, $\hat\Psi^{\dag}({\bm x})$ satisfy the canonical anti-commutation 
rules $\{\hat\Psi({\bm x}),\hat\Psi^{\dag}({\bm y})\} = \delta({\bm x}-{\bm y})$. 
For brevity, we used the convention that ${\bm x}\equiv\{x_1,\dots, x_d\}$ and, therefore, 
$\delta({\bm x}-{\bm y}) \equiv \prod_{i=1}^{d}\delta(x_{i}-y_{i})$.
Thanks to this fact, the eigenfunctions of the one-particle differential operator $-\sum_{i=1}^{d}\partial^{2}_{x_i}$
are easily found
\be
\Phi_{{\bm p}}({\bm x}) = \prod_{i=1}^{d}\varphi_{p_{n_i}}(x_i),
\ee
with ${\bm p}\equiv\{p_{n_1},p_{n_2},\dots,p_{n_d}\}$ and
\be
\varphi_{p_{n_i}}(x_i) = \sqrt{\frac{2}{L}}\sin \left[ p_{n_i} \left(x_i+ \frac{L}{2}\right) \right],\; p_{n_i} = \frac{\pi}{L} n_i,\\
\ee
with $n_i\in\mathbb{N}$. By using the previous eigenfunctions, the free-fermionic 
operators are easily introduced
\be
\hat \eta({\bm p}) = \int_{\mathcal{V}}  \prod_{i=1}^{d} dx_i \Phi_{{\bm p}}({\bm x}) \hat \Psi({\bm x}),
\ee
and the Hamiltonian (\ref{H}) is readily diagonalized
\be\label{H_diag}
\hat H = \sum_{n_1,\dots,n_d=0}^{\infty} \sum_{i=1}^{d}\frac{ p_{n_i}^{2}}{2m} \,\hat{n}({\bm p}),
\ee
with  $\hat{n}({\bm p})\equiv \hat\eta^{\dag}({\bm p})\hat\eta({\bm p})$ 
the fermionic mode occupation operator.
Notice that the Hamiltonian commutes with the total number of particles operator
\be
\hat{N} = \int_{\mathcal{V}} \prod_{i=1}^{d}dx_i \, \hat\Psi^{\dag}({\bm x})\hat\Psi({\bm x})
 = \sum_{n_1,\dots,n_d=0}^{\infty}\hat{n}({\bm p}),
\ee
therefore, $\hat H$ and $\hat N$ can be simultaneously diagonalized in the many-body Hilbert space. 
Since the excitation spectrum of the Hamiltonian is non-negative, the ground 
state without fixing the number of particles is the vacuum state $|0\rangle$, 
such that $\hat\eta({\bm p})|0\rangle = 0,\;\forall {\bm p}$.

At time $t=0$ the system is divided into two halves along the $x_1$ coordinate, 
namely $\mathcal{L}$ ($x_1<0$) and $\mathcal{R}$ ($x_1>0$). 
Otherwise, the other directions are left unchanged during the quench.
The two subsystems are initially uncorrelated and the total initial Hamiltonian is the direct sum
of two independent Hamiltonians which can be easily diagonalized following the same approach 
used so far for the post-quench Hamiltonian.
The only difference appears along the $x_1$ direction due to the new boundary conditions at $x_1=0$. 
Therefore, the normalized eigenfunctions, building-up the one-particle Hilbert space of the semi-domains
i.e. $[-L/2,0]\otimes \mathcal{V}_{d-1}$ or $[0,L/2]\otimes \mathcal{V}_{d-1}$, are given by
\be
\Phi^{\pm}_{{\bm q}}({\bm x}) = \phi^{\pm}_{q_{m_1}}(x_1) \prod_{i=2}^{d}\varphi_{p_{m_i}}(x_i),
\ee
where ${\bm q}\equiv\{q_{m_1},p_{m_2},\dots,p_{m_d}\}$ and
\be
\phi^{\pm}_{q_{m_1}}(x_1) = \frac{2}{\sqrt{L}}\sin\left( q_{m_1} x_1 \right) \theta(\pm x_1),\; q_{m_1} = \frac{2\pi}{L} m_1,
\ee
with $m_1\in\mathbb{N}$ and where $\theta(x)$ is the Heaviside function such 
that $\theta(x)=1$ for $x>0$, and zero otherwise.
Using these functions the half-domain Hamiltonians are straightforwardly diagonalized. 
For example, in the sub-domain $[0,L/2]\otimes \mathcal{V}_{d-1}$ one has
\bea\label{H_0}
\hat H^{+}_{0} & = & \int_0^{L/2}dx_1\int_{\mathcal{V}_{d-1}} \prod_{i=2}^{d}dx_i \,\hat\Psi^{\dag}({\bm x}) 
\left\{ - \frac{1}{2m}\sum_{i=1}^{d}\partial^{2}_{x_i} \right\} \hat\Psi({\bm x})\nonumber\\
& = & \sum_{m_1,\dots,m_d=0}^{\infty} \sum_{i=1}^{d} \frac{q_{m_i}^{2}}{2m} \, \hat\xi^{\dag}({\bm q})\hat\xi({\bm q}),
\eea
where the fermionic fields $\hat\Psi({\bm x})$, $\hat\Psi^{\dag}({\bm x})$ are related to the diagonal operators via
\bea
\hat\Psi({\bm x}) = \sum_{m_1,\dots,m_d=0}^{\infty} \Phi^{+}_{{\bm q}}({\bm x}) \hat\xi({\bm q}),\quad x_1>0.
\eea
Similar arguments are valid for the negative half-domain. 

Finally, in order to characterize the post-quench dynamics, it is useful to know the decomposition of the
initial one-particle eigenfunctions in terms of the the post-quench basis, i.e.
\be\label{phi_to_varphi}
\Phi^{\pm}_{{\bm q}}({\bm x}) = \sum_{{\bm p}=0}^{\infty} \mathbb{A}^{\pm}_{{\bm p},{\bm q}} \Phi_{{\bm p}}({\bm x}),
\ee 
where, in order to simplify the notation, the sum runs over the indeces $n_{i}$ defining the vector ${\bm p}$,
and the overlap $\mathbb{A}^{\pm}_{{\bm p},{\bm q}}$ implicitly depends on all the components of 
the momenta ${\bm p}$ and ${\bm q}$.
Notice in particular that, thanks to the specific geometry of the problem, the overlap reduces to the only
diagonal part in all momenta apart from $p_{1}=\pi n_1 /L$ and $q_1=2\pi m_1 /L$, 
i.e. the quenched momenta:
\be
\mathbb{A}^{\pm}_{{\bm p},{\bm q}} =  A^{\pm}_{n_1,m_1}\prod_{i=2}^{d}\delta_{n_i,m_i},
\ee
where $A^{\pm}_{n,m}$ has been explicitly evaluated in Ref. \cite{ck14}, 
obtaining
\be
A^{\pm}_{n,m} = \left\{\begin{array}{cc}
\pm \frac{4\sqrt{2} m \sin(n \pi /2) }{\pi(4m^2-n^2)} & \mathrm{for}\; n\;\mathrm{odd}, \\
 & \\
 \frac{(-1)^m}{\sqrt{2}}\delta_{n,2m} & \mathrm{for}\; n\;\mathrm{even}.
\end{array}\right. 
\ee

\section{Observables}\label{Obs}
We are interested in the non-equilibrium stationary behavior 
of the system after a local quantum quench along the spatial direction $x_1$. 

It is worth mentioning the fact that, the quench protocol we are considering is ``local''
in the sense that the post-quench dynamics is induced by a local coupling. 
However, unlike the zero-temperature local quantum quenches 
\cite{localCFT,localentropy,peschel05,entr_cc13},
in our setup the excess energy density in each of the two halves 
after the quench is finite, making such a protocol
similar to a ``geometric'' quench \cite{mpc10,ah14}.

Moreover, in order to fully characterize the non-equilibrium properties, it is natural to look 
at the energy and particle transport along such direction, 
after having integrated out the remaining coordinates. 
Indeed, in the TD limit and for large times, all currents 
flowing in the directions orthogonal to $x_1$ will be na\"ively vanishing. 

Therefore, it is natural to introduce the linear density of particle and the linear density 
of energy along $x_1$, as
\bea
n(x_1,t) & \equiv & \frac{1}{L^{d-1}}\int_{\mathcal{V}_{d-1}} \prod_{i=2}^{d} dx_i 
\langle \hat\Psi^{\dag}({\bm x}) \hat\Psi({\bm x})\rangle_{t},\\
\mathcal{E}(x_1,t) & \equiv & \frac{1}{L^{d-1}}\int_{\mathcal{V}_{d-1}} \prod_{i=2}^{d} dx_i 
\langle \hat\Psi^{\dag}({\bm x}) \mathcal{H}({\bm x})\hat\Psi({\bm x})\rangle_{t},
\eea
with the one-particle Hamiltonian's differential operator 
$\mathcal{H}({\bm x})\equiv -\sum_{i=1}^{d}\partial^{2}_{x_i}/(2m)$. 
These quantities will be useful in order to evaluate the particle and energy currents. 
In the following, let us specialize the discussion to the particle current 
$\mathbf{J}\equiv \{J_{x_1},\dots,J_{x_d}\}$, the same arguments will be valid for the energy current. 

Thus, the continuity equation for the current of particles 
in $d$ dimensions reads 
$\sum_{i=1}^{d}\partial_{x_i} J_{x_i}({\bm x},t) = -\partial_{t} \langle \hat\Psi^{\dag}({\bm x}) \hat\Psi({\bm x})\rangle_{t}$.
Integrating such equation in $\mathcal{V}_{d-1}$ and considering the fact that, in the TD limit, 
the system is in equilibrium along all the directions orthogonal to $x_1$, one immediately has
\be
\partial_{x_1}\frac{1}{L^{d-1}}\int_{\mathcal{V}_{d-1}}\prod_{i=2}^{d} dx_i  J_{x_1}({\bm x},t) = -\partial_{t}n(x_1,t),
\ee
from which we have the average current of particles flowing along the quenched direction
\be\label{particle_current}
\mathcal{J}(x_1,t) = -\int_{-\infty}^{x_1} dz\,\partial_{t} n(z,t),
\ee
where the integration boundaries are opportunely fixed in order to have vanishing current at 
$x_1=\pm\infty$. It is straightforward to show the equivalence 
$\mathcal{J}(x_1,t) \equiv 2\,{\rm Im} \left[ L^{1-d} \int_{\mathcal{V}_{d-1}} \prod_{i=2}^{d} dx_i 
\langle \hat\Psi^{\dag}({\bm x})\partial_{x_1} \hat\Psi({\bm x}) \rangle_{t}\right]$.
Following the same lines, we introduce the average current of energy along $x_1$
\be\label{energy_current}
\vartheta(x_1,t) = -\int_{-\infty}^{x_1} dz\,\partial_{t} \mathcal{E}(z,t),
\ee
which exactly agrees with the definition

$
\vartheta(x_1,t) \equiv 2\,{\rm Im} \left[ L^{1-d} \int_{\mathcal{V}_{d-1}} \prod_{i=2}^{d} dx_i 
\langle [\partial_{x_1}\hat\Psi^{\dag}({\bm x})]\mathcal{H}({\bm x})\hat\Psi({\bm x}) \rangle_{t}\right].
$

\section{Post-quench dynamics}\label{PQD}
The dynamics is unitarily generated by the Hamiltonian (\ref{H}) starting from an out-of-equilibrium 
initial state constructed as a tensor product of two thermal density matrices at two different 
temperatures and chemical potential, i.e. the ``Grand Canonical tensor state'' 
$\hat\rho_{0} = \hat\varrho_{-}(\mu_{\mathcal{L}},\beta_{\mathcal{L}})
\otimes\hat\varrho_{+}(\mu_{\mathcal{R}},\beta_{\mathcal{R}})$, 
where $\hat\varrho_{\pm}(\mu,\beta) = Z^{-1} \exp [ - \beta(\hat H_{0}^{\pm}-\mu \hat N^{\pm})]$.
The state $\hat\varrho_{\pm}(\mu,\beta)$ is constructed in such a way that, whenever one considers the 
small temperature behavior, i.e. $\beta_{\mathcal{L/R}}\gg 1$, the initial density matrix $\hat\rho_{0}$ 
reduces to the tensor product of the two filled Fermi seas 
$|N_{\mathcal{L}}\rangle \langle N_{\mathcal{L}} | \otimes 
|N_{\mathcal{R}}\rangle \langle N_{\mathcal{R}} |$, 
where $|N_{\mathcal{L/R}}\rangle = \prod_{k}^{k^{*}_{\mathcal{L/R}}} 
\hat\eta^{\dag}_{\mathcal{L/R}}(k)|0\rangle$,
with the Fermi momentum $k^{*}_{\mathcal{L/R}}$ being an implicit function of the 
chemical potential $\mu_{\mathcal{L/R}}$.

As a consequence of the factorization, the two spatial regions ($\mathcal{L}$ and $\mathcal{R}$) 
are initially uncorrelated. Furthermore, such a state is quadratic in the local field operators and, 
therefore, the Wick's theorem applies.
Moreover, since the Hamiltonian is quadratic in the fermionic operators, 
the evolved state keeps its gaussian character and the Wick's theorem still applies. 
Consequently, all observables are derived from the two-point correlation function. 
Therefore, we focus our attention on the time-evolution of the two-point correlation 
function which is given by
\be\label{Corr_t}
C({\bm x},{\bm y};t) = \sum_{{\bm p},{\bm q}=0}^{\infty} \Phi_{{\bm p}}({\bm x})^{*}\Phi_{{\bm q}}({\bm y}) 
\langle \hat\eta^{\dag}({\bm p}) \hat\eta({\bm q}) \rangle_{t}, 
\ee
where, once again, in order to simplify the notation, the sum implicitly runs over 
all the components of the vectors ${\bm p}$ and ${\bm q}$. 
Following Ref. \cite{ck14}, one can straightforwardly rewrite 
the time evolved correlation function as
\be\label{Corr_t_v2}
C({\bm x},{\bm y};t)  =  \sum_{{\bm q}=0}^{\infty} \Bigg[ \frac{\Phi^{-}_{{\bm q}}({\bm x},t)^{*}\Phi^{-}_{{\bm q}}({\bm y},t)}{1+\exp[\beta_{\mathcal{L}}(E_{\bm q}-\mu_{\mathcal{L}})]}
+  \frac{\Phi^{+}_{{\bm q}}({\bm x},t)^{*}\Phi^{+}_{{\bm q}}({\bm y},t)}{1+\exp[\beta_{\mathcal{R}}(E_{\bm q}-\mu_{\mathcal{R}})]}  \Bigg],
\ee
where, as usual ${\bm q}=\{q_{m_1},p_{m_2},\dots,p_{m_d}\}$, $E_{\bm q}\equiv {\bm q}^2/(2m)$ 
and we introduced the time-evolved one-particle eigenfunctions (neglecting an overall phase factor)
\bea\label{Phi_x_t}
\Phi^{\pm}_{{\bm q}}({\bm x},t) & = & \sum_{{\bm p}=0}^{\infty} \mathbb{A}^{\pm}_{{\bm p},{\bm q}} 
\Phi_{{\bm p}}({\bm x}) {\rm e}^{-i E_{\bm p} t} \\
& = & \prod_{i=2}^{d} \varphi_{p_{m_i}}(x_i) \sum_{n_1=0}^{\infty}A^{\pm}_{n_1,m_1}\varphi_{p_{n_1}}(x_1) \mathrm{e}^{-i t p^{2}_{n_1}/(2m)}.\nonumber
\eea

\subsection{The one-dimensional equivalence}
Once we have a closed expression for the time-dependent two-point 
correlation function in the entire domain $\mathcal{V}$,
the next step consists in integrating out the coordinates orthogonal to the quenched one.
Once again, let us start with the particle density. In this case, by evaluating 
the correlation function in Eq. (\ref{Corr_t_v2}) 
at the same spatial point ${\bm x}={\bm y}$, after integrating over $\mathcal{V}_{d-1}$, 
and using the orthogonality of the one-particle eigenfunctions, 
we can reduce the original $d$-dimensional problem, 
to an equivalent $1D$-problem. Indeed, the linear density of particles becomes, in the TD limit,
\be\label{n_x_t_1D}
n(x,t) = \sum_{l=0}^{\infty} \left[ |\phi^{-}_{q_l}(x,t)|^2 \tilde n_{\mathcal{L}}(q_l)  
+  |\phi^{+}_{q_l}(x,t)|^2 \tilde n_{\mathcal{R}}(q_l)  \right],
\ee
where $\tilde n_{\mathcal{L/R}}(q)$ plays the role of an affective 
one-dimensional mode occupation 
\be
\tilde n_{\mathcal{L/R}}(q) =\int_{0}^{\infty}\prod_{i=2}^{d} \frac{dp_i}{\pi^{d-1}} 
\frac{1}{1+\exp[\beta_{\mathcal{L/R}}(E_{{\bm p}}-\mu_{\mathcal{L/R}})]},
\ee
with ${\bm p} = \{p_1,\dots,p_d\}$ and $q\equiv p_1$.

Similar arguments are valid for the energy density. In other words, also
in this case, the original problem can be mapped to an analogous one-dimensional problem. 
However, in this case, similarly to what happens in the genuinely $1D$ situation, 
the spatial derivative along the quenched direction, i.e. $-\partial^{2}_{x_1}$, 
when applied on the Eq. (\ref{Phi_x_t}), 
brings an extra dependence on the quenched momenta $p_{n_1}$ inside the sum, which obviously 
changes the linear combination of the time-evolved one-particle eigenfunction in Eq.(\ref{Phi_x_t}). 
In order to avoid such a mixing, we proceed in the following way:
we explicitly evaluate the spatial derivative along the all directions orthogonal to 
$x_1$ (this can be done because the eigenfunctions in such directions do not evolve at all), 
otherwise we still treat the quenched momentum as an ordinary differential operator. 
In this way, the problem can be exactly casted in an equivalent one-dimensional problem
\bea\label{E_x_t_1D}
\mathcal{E}(x,t)  & = & \sum_{l=0}^{\infty} \Bigg\{  \phi^{-}_{q_l}(x,t)^{*}\left[\tilde\mathcal{E}_{\mathcal{L}}(q_l)- \tilde n_{\mathcal{L}}(q_l)
\frac{\partial^{2}_{x}}{2m} \right] \phi^{-}_{q_l}(x,t)\nonumber\\
& + & \phi^{+}_{q_l}(x,t)^{*} \left[\tilde\mathcal{E}_{\mathcal{R}}(q_l)- \tilde n_{\mathcal{R}}(q_l)\frac{\partial^{2}_{x}}{2m}\right] \phi^{+}_{q_l}(x,t)\Bigg\},
\eea
where we introduced an effective one-dimensional energy distribution function
\be\label{1d_energy_function}
\tilde\mathcal{E}_{\mathcal{L/R}}(q) = \int_{0}^{\infty}\prod_{i=2}^{d} \frac{dp_i}{\pi^{d-1}} 
\frac{\sum_{i=2}^{d} p^2_{i}/(2m)}{1+\exp[\beta_{\mathcal{L/R}}(E_{{\bm p}}-\mu_{\mathcal{L/R}})]}.
\ee
Notice how, in this case, the total effective one-dimensional energy operator is more involved, being 
composed by an algebraic term $\tilde\mathcal{E}_{\mathcal{L/R}}(q)$ plus a differential operator 
$\tilde n_{\mathcal{L/R}}(q)\partial^{2}_{x}/(2m)$. In particular, whenever $d=1$, the numerator in Eq. (\ref{1d_energy_function})
becomes identically vanishing and the total operator reduces to 
$-[\partial^{2}_{x}/(2m)]/\{1 + \exp[\beta_{\mathcal{L/R}}(q^2/(2m)-\mu_{\mathcal{L/R}})] \}$ as expected.

The effective functions describing the equivalent one-dimensional 
problem are explicitly evaluated in \ref{app_effective_functions}.

\begin{figure}[t!]
\center\includegraphics[width=0.33\textwidth]{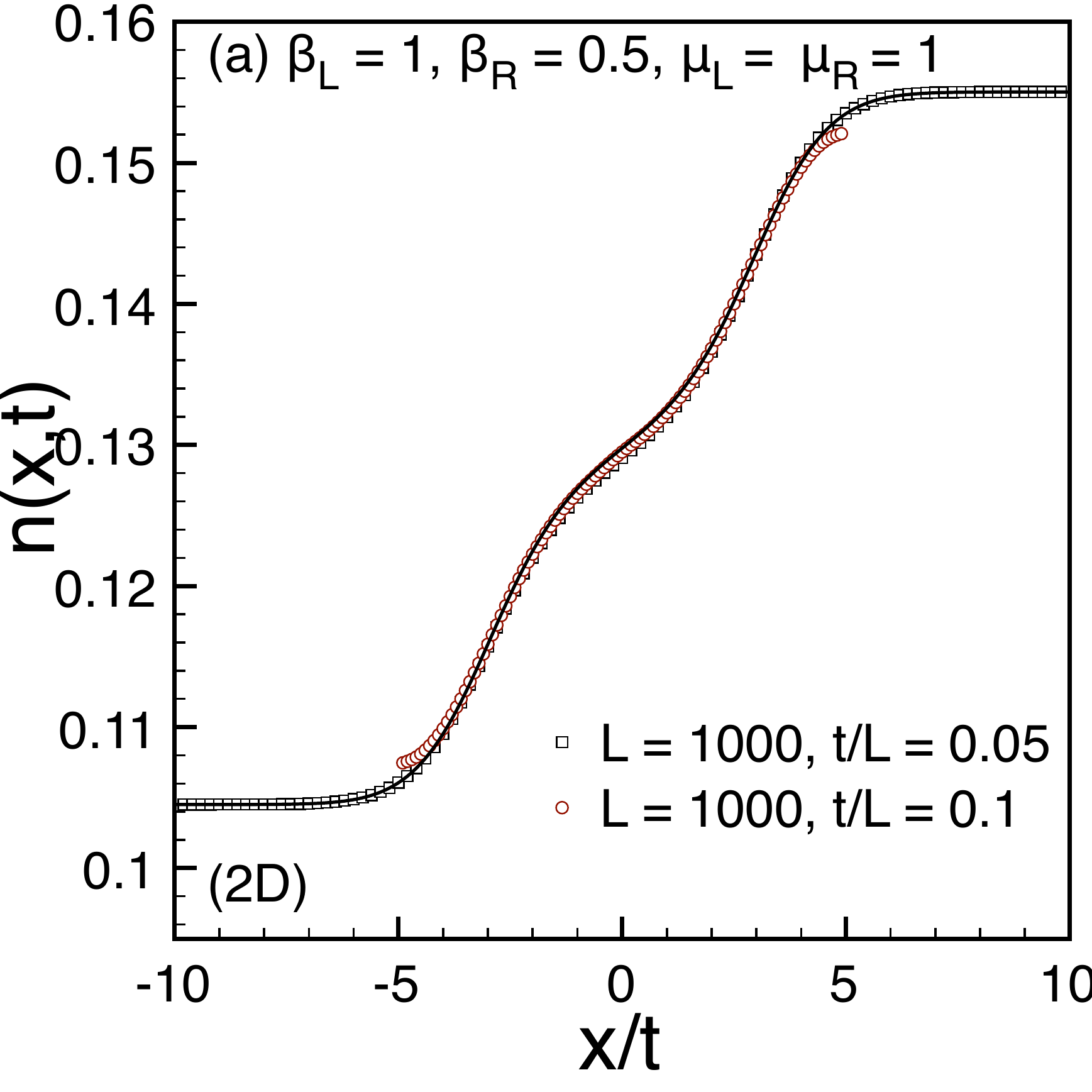}\includegraphics[width=0.33\textwidth]{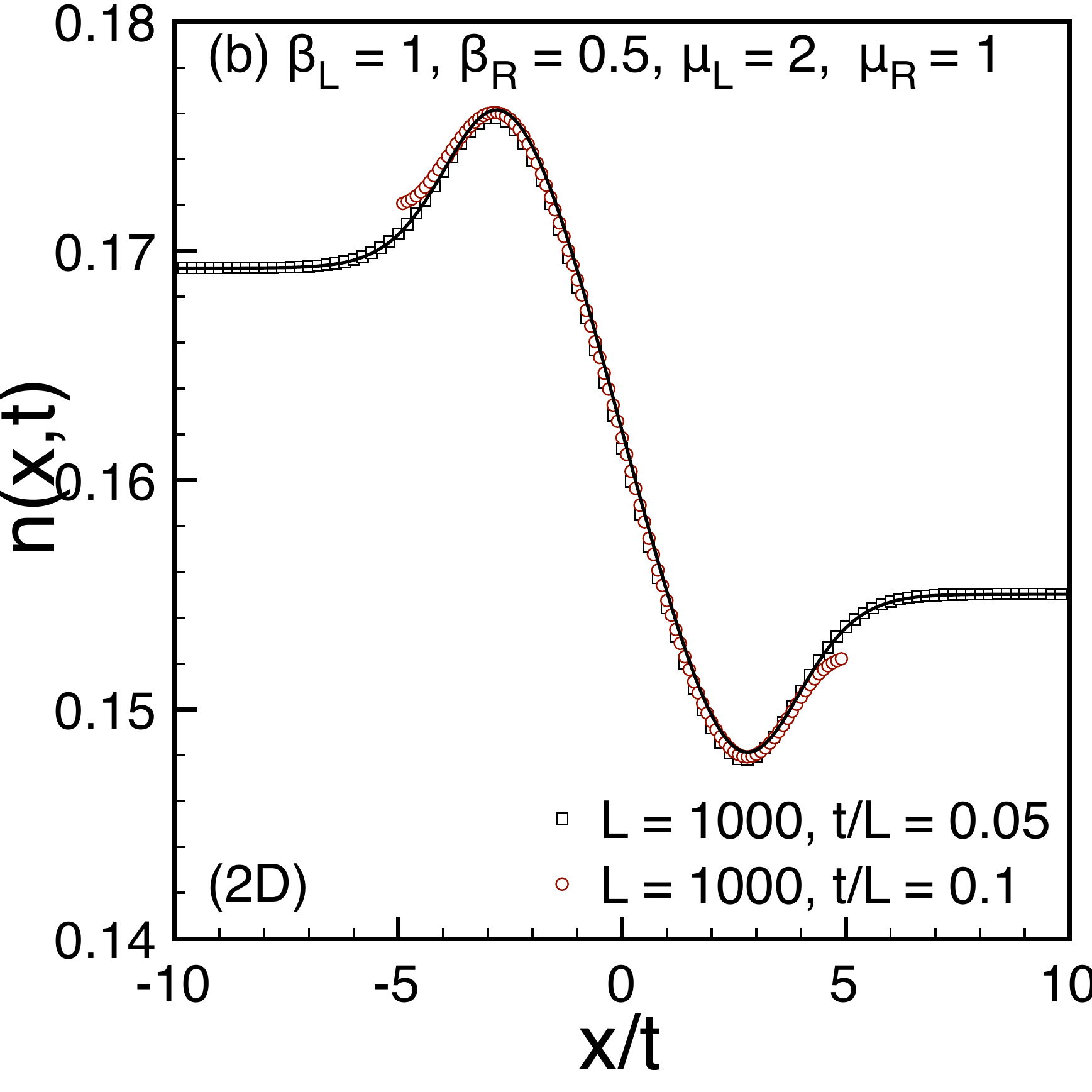}\includegraphics[width=0.33\textwidth]{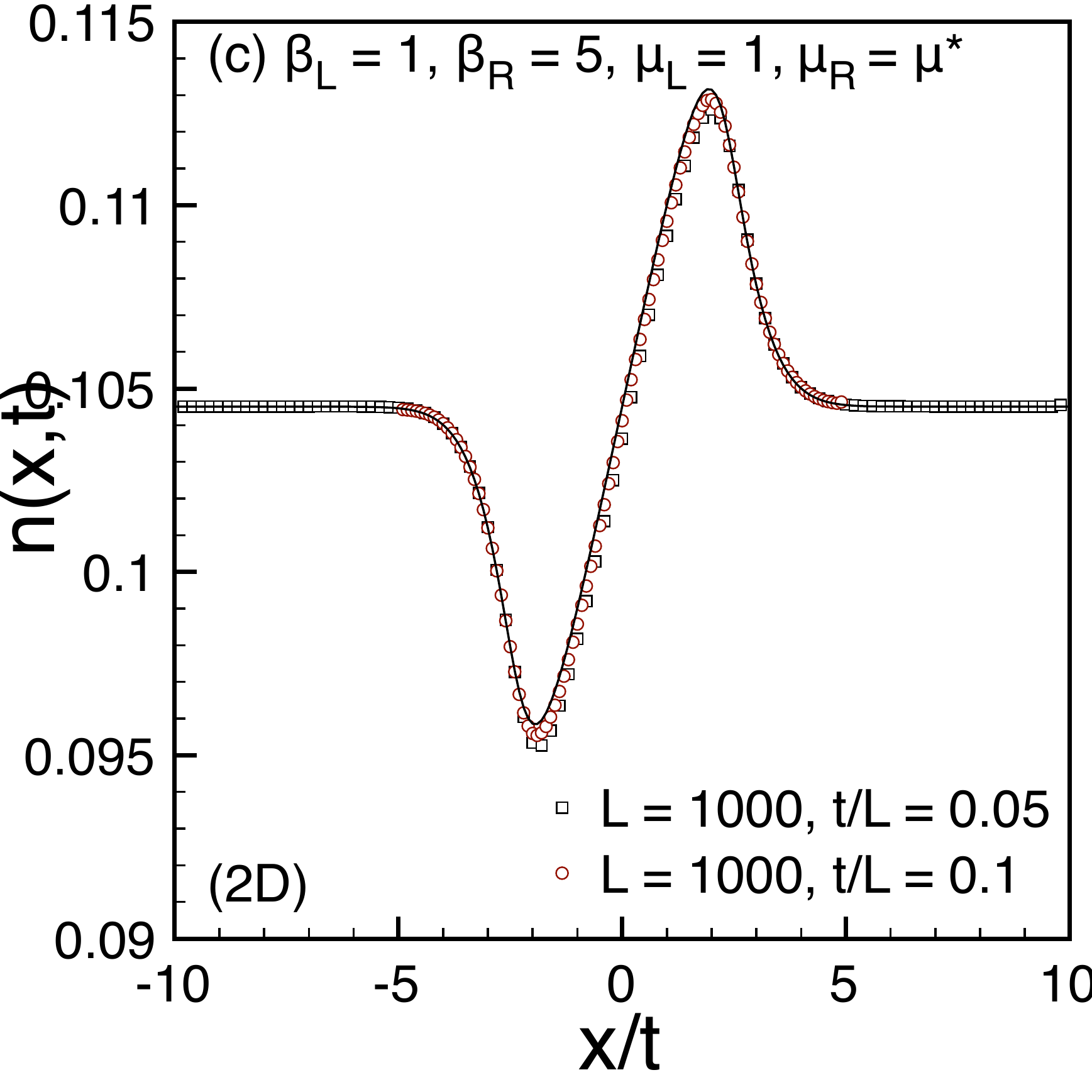}\\
\center\includegraphics[width=0.33\textwidth]{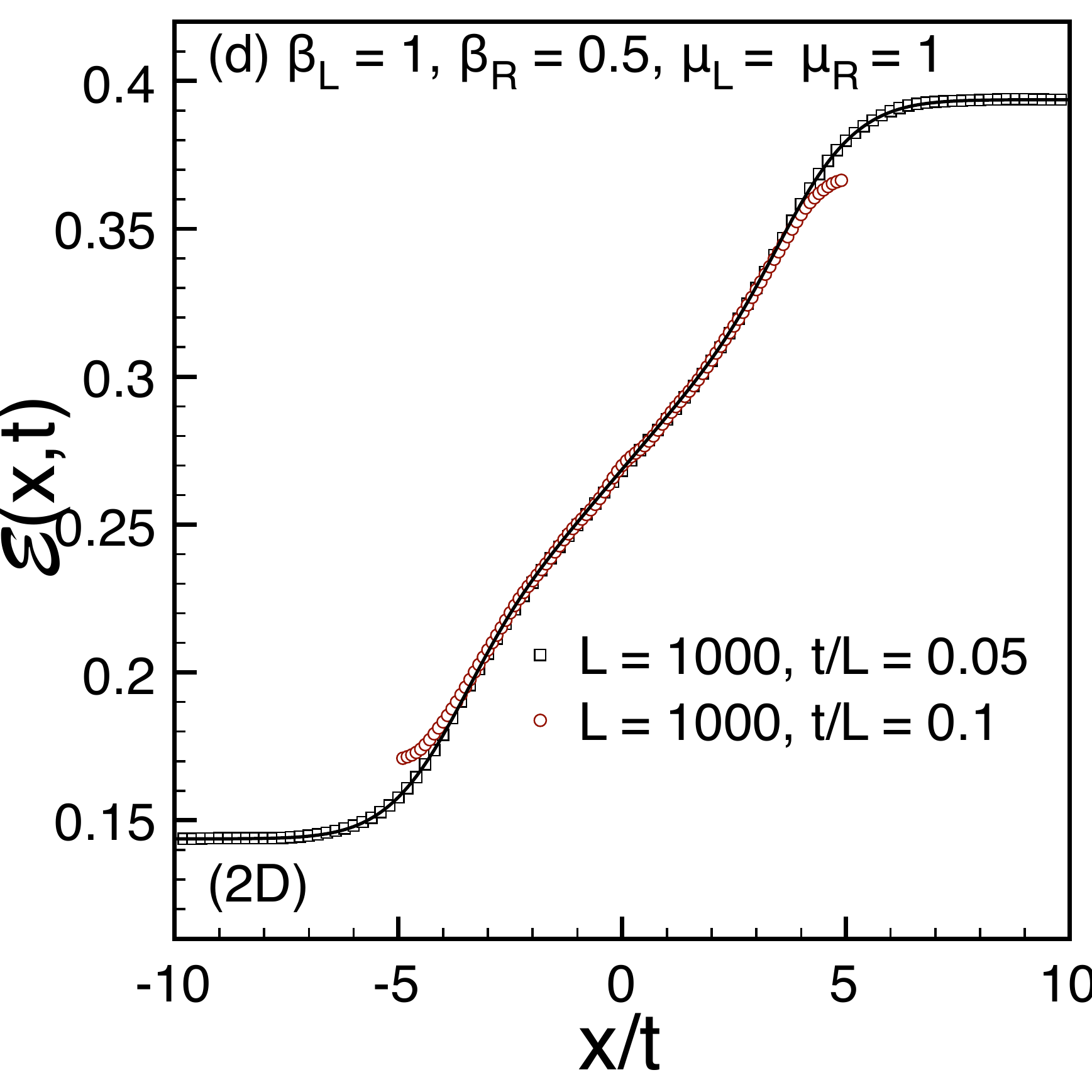}\includegraphics[width=0.33\textwidth]{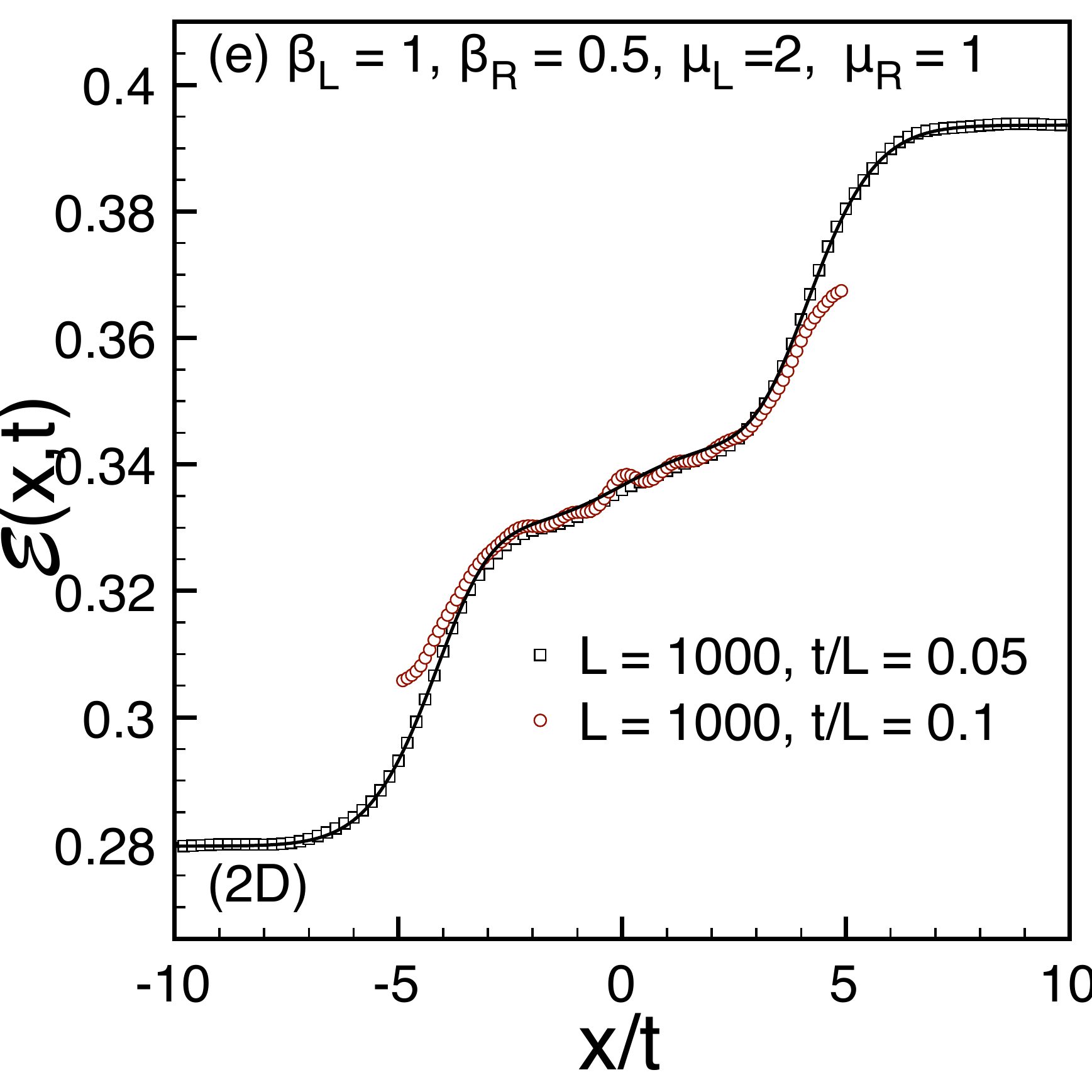}\includegraphics[width=0.33\textwidth]{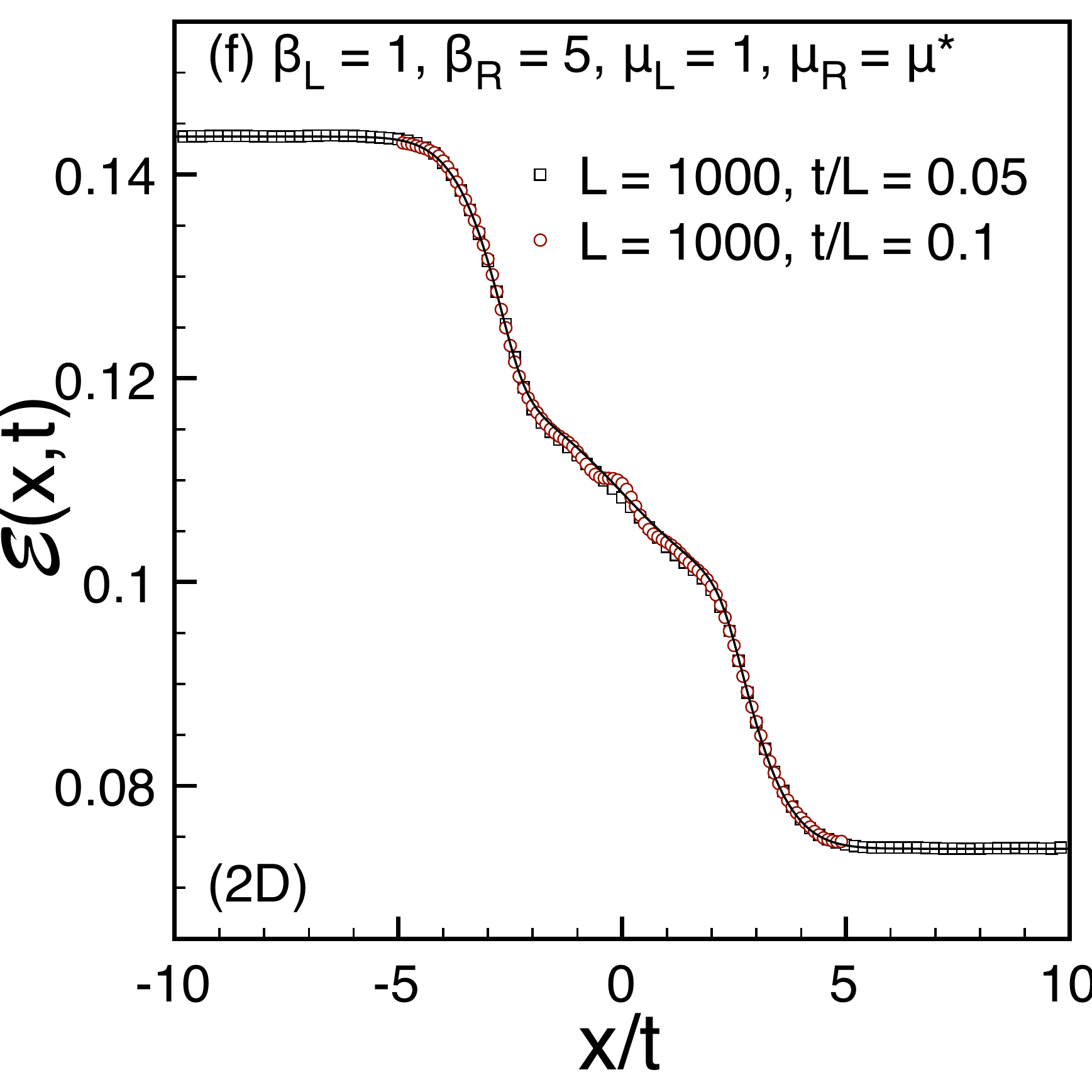}
\caption{(a,b,c) Numerically evaluated particle density profile via Eq. (\ref{n_x_t_1D}) in two dimensions
vs the scaling variable $x/t$ for different rescaled times $t/L$ (symbols). 
The full lines are the analytic result (\ref{nxt_scaling}). (d,e,f) Numerically
 evaluated energy density profile via Eq. (\ref{E_x_t_1D}) in two dimensions 
 vs the scaling variable $x/t$ for different sizes $L$
 and rescaled times $t/L$ (symbols). The full lines are the analytic result (\ref{Ext_scaling}). 
 In the rightmost column, i. e. (c,f), $\mu_{\mathcal{R}}=\mu^{*}=1.31298$ in order to prepare
 initially the systems with the same particle densities $n_{\mathcal{L/R}}=0.104506$.
 For all data the fermions' mass has been fixed to $m=1/2$.} 
\label{fig_nE_2D_scaling}
\end{figure}

\begin{figure}[t!]
\center\includegraphics[width=0.33\textwidth]{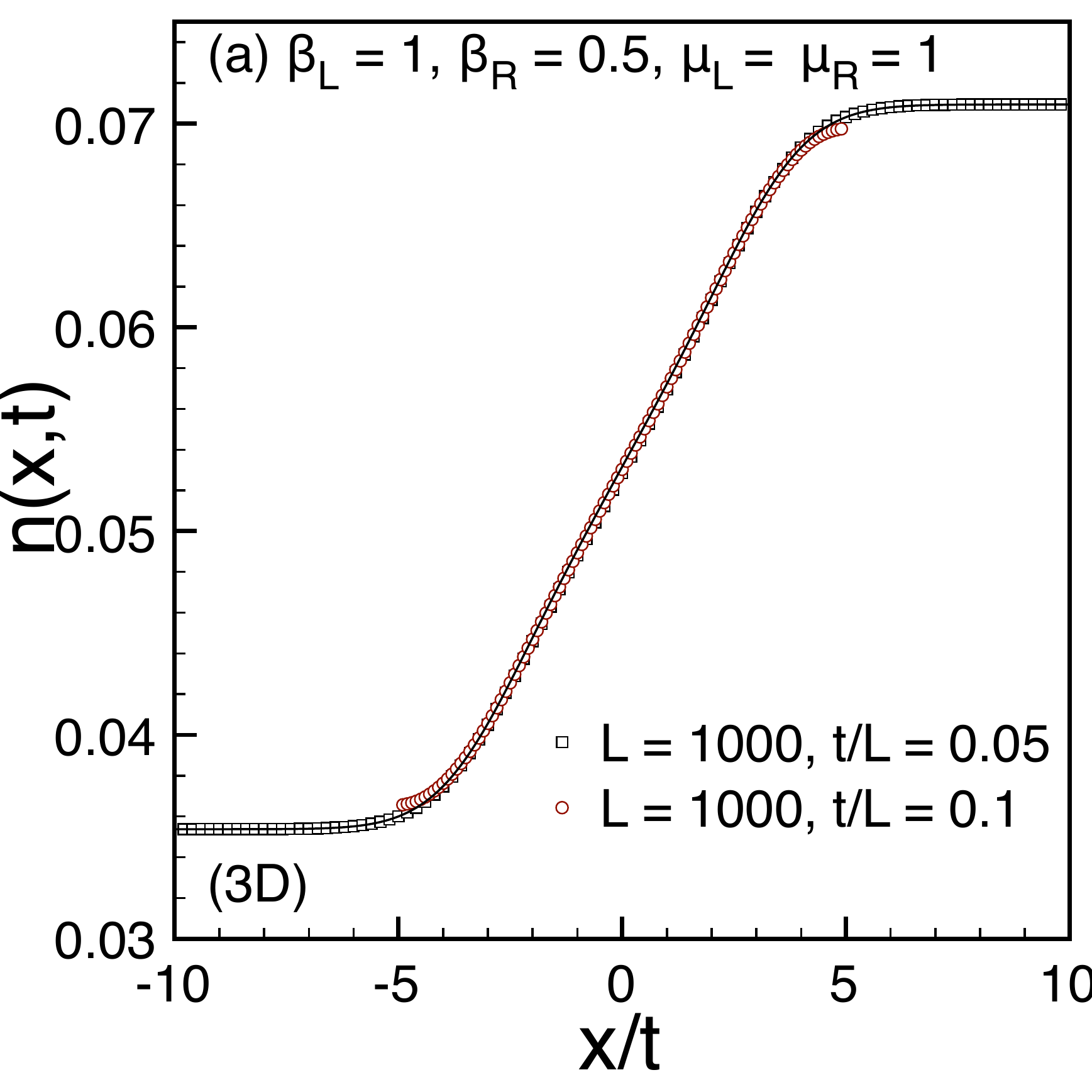}\includegraphics[width=0.33\textwidth]{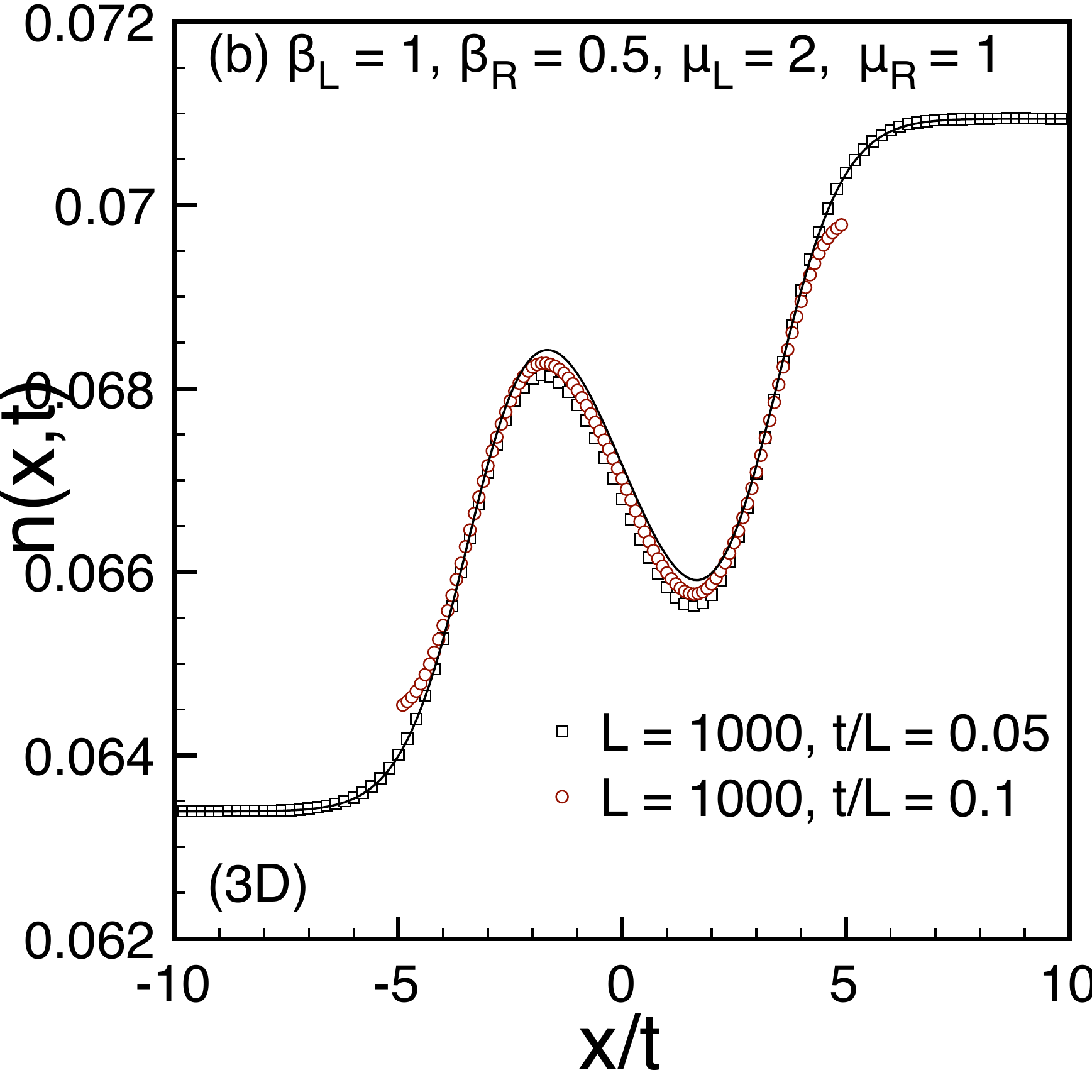}\includegraphics[width=0.33\textwidth]{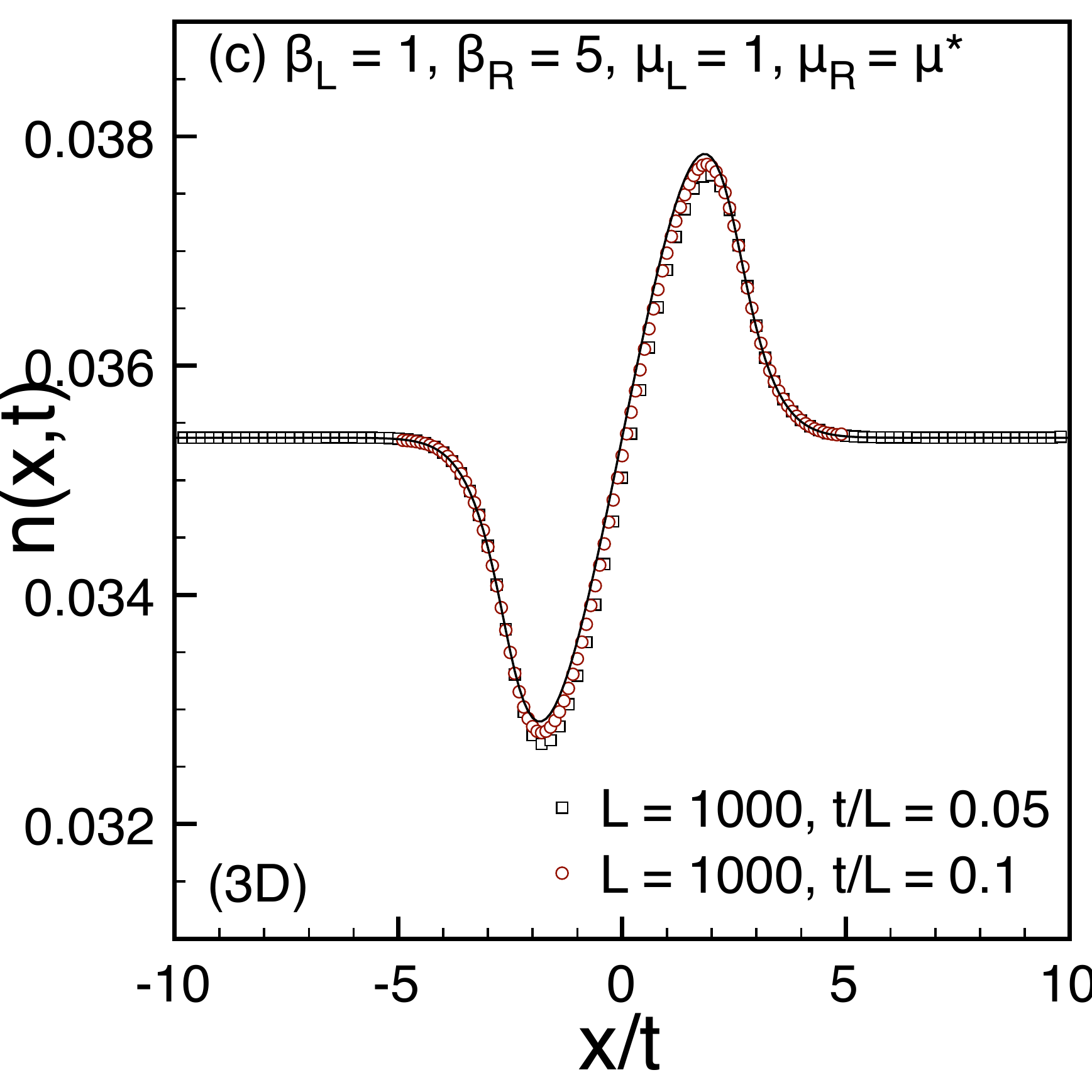}\\
\center\includegraphics[width=0.33\textwidth]{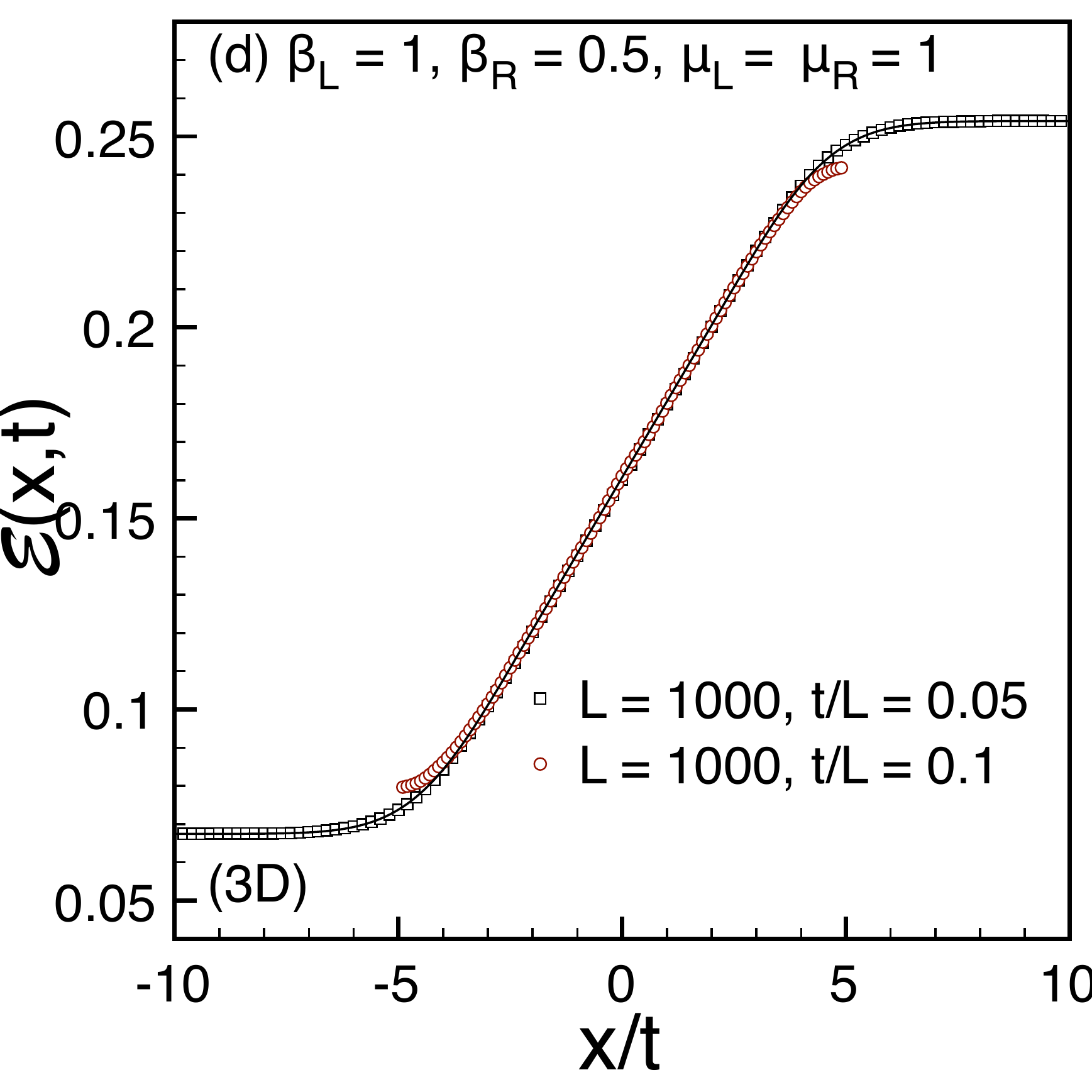}\includegraphics[width=0.33\textwidth]{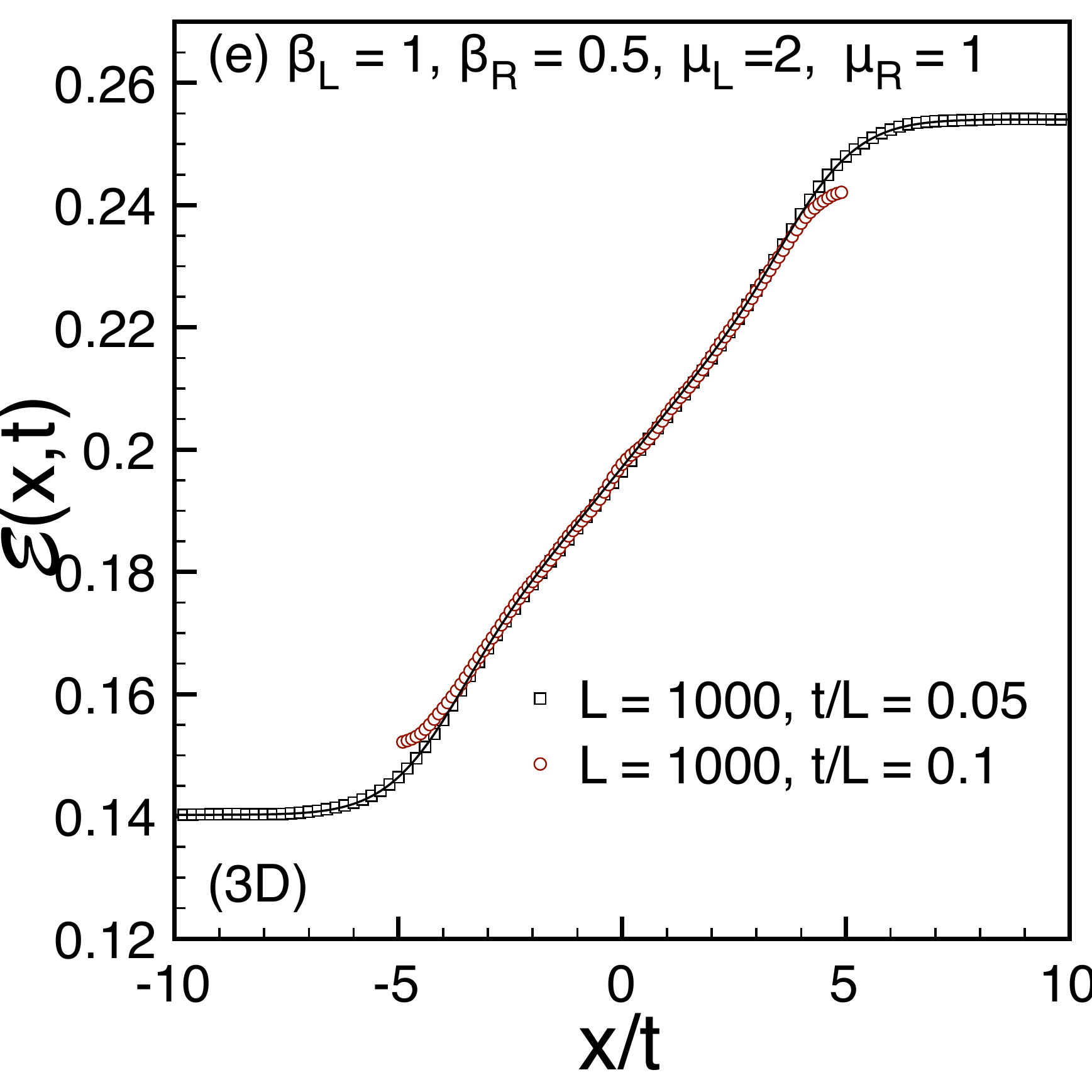}\includegraphics[width=0.33\textwidth]{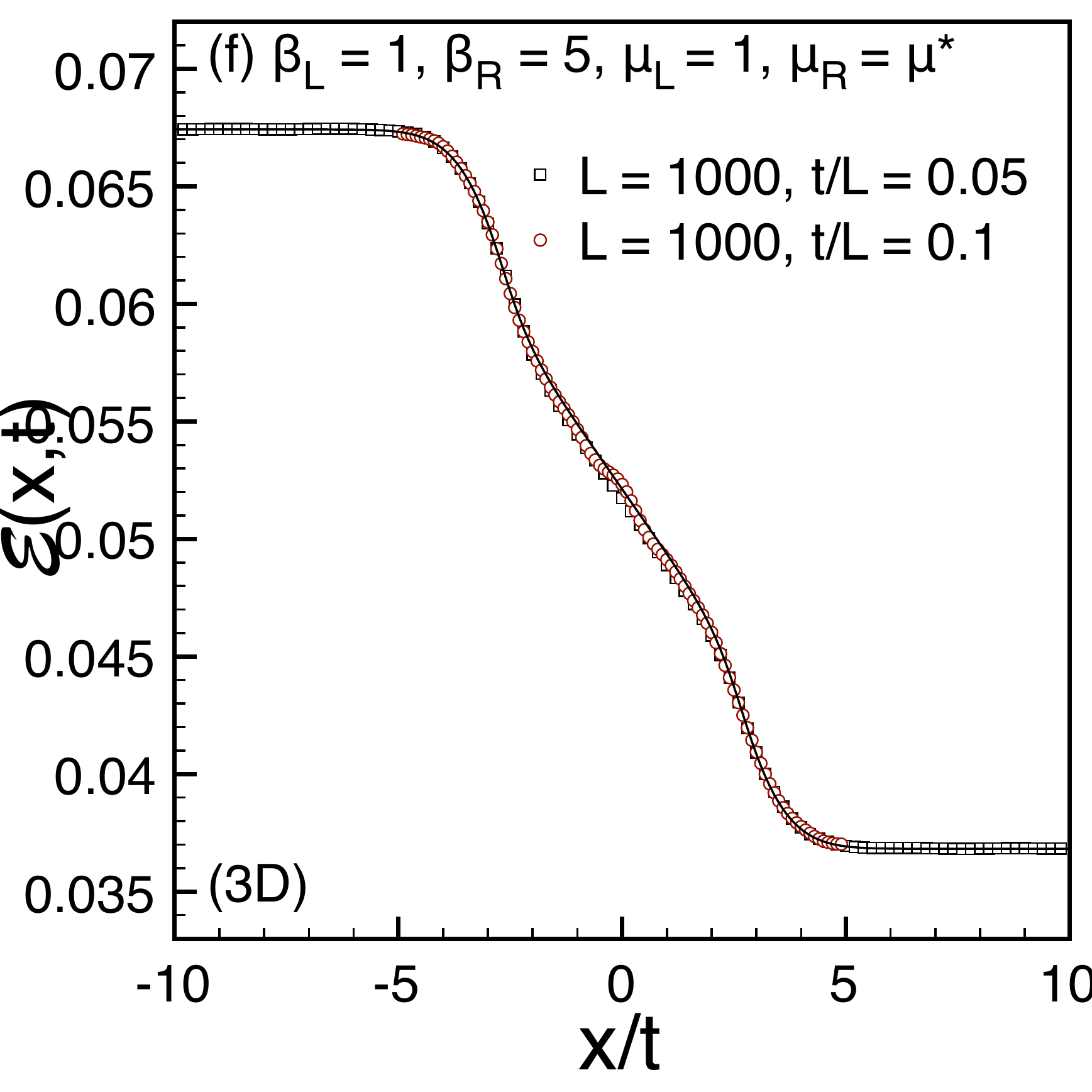}
\caption{(a,b,c) Numerically evaluated particle density profile via Eq. (\ref{n_x_t_1D}) in three dimensions
vs the scaling variable $x/t$ for different rescaled times $t/L$ (symbols). 
The full lines are the analytic result (\ref{nxt_scaling}). (d,e,f) Numerically
 evaluated energy density profile via Eq. (\ref{E_x_t_1D}) in three dimensions 
 vs the scaling variable $x/t$ for different sizes $L$
 and rescaled times $t/L$ (symbols). The full lines are the analytic result (\ref{Ext_scaling}). 
 In the rightmost column, i. e. (c,f), $\mu_{\mathcal{R}}=\mu^{*}=1.61644$ in order to prepare
 initially the systems with the same density particles $n_{\mathcal{L/R}}=0.03537$.
 For all data the fermions' mass has been fixed to $m=1/2$. } 
\label{fig_nE_3D_scaling}
\end{figure}

 \subsection{Semiclassical description of the time-dependent profile of local 
 observables along the quenched direction}
 Since the original $d$-dimensional problem can be easily mapped to an equivalent $1D$ problem,
 we can straightforwardly apply the same semiclassical approach proposed in Ref. \cite{ck14}.
 Also in this general case, the dynamics of the particle-density profile as well as of the energy-density 
 profile along the quenched direction can be fully characterized, in the TD limit, 
 by a semiclassical description in the $2$-dimensional phase-space \cite{cark12,wck13}. 
 Once again, we stress that such a construction can not reproduce the dynamics of the correlations.
 
Let us start considering, for example, the unidimensional particle density. 
Indeed, from the previous paragraph we already know 
that the dynamics of the linear density of particles is fully characterized by a 
specific one-dimensional distribution associated to
the equivalent $1D$ problem. 
Therefore, we can associate to each phase-space point a local initial packet of particles 
$n_{0}(x,p) dp dx$. Then, each of them evolves following a classical trajectory 
 $x^{\pm}(t) = x_{0} \pm v_{p} t$, with velocity $v_{p} \equiv \partial_{p} E_p = p/m$, 
 where the sign refers to  the left($-$) and right($+$)
 movers, similarly to what was already done in Ref. \cite{ck14}. 
 Finally, in our case, the initial particle distribution in the phase-space 
 (associated to the quenched degrees of freedom) is
 \be\label{n0_px}
 n_{0}(x,p) = \frac{\tilde n_{\mathcal{L}}(p)}{\pi}\theta(-x)+ \frac{\tilde n_{\mathcal{R}}(p)}{\pi}\theta(x),
 \ee
 from which we can straightforwardly obtain the time-evolved density profile
 \be
 n(x,t) =\frac{1}{2} \sum_{\sigma=\pm 1} \int dp\,\int dx_0 \, n_{0}(x_0,p)\delta(x-x_0-\sigma pt/m).
 \ee
 Injecting Eq.(\ref{n0_px}) in the latter equation, we finally obtain
 \be\label{nxt_scaling}
 n(x,t) = \frac{1}{2}\left\{\begin{array}{cc}  n_{\mathcal{R}} + f_{\mathcal{L}}(mx/t,\infty)+ f_{\mathcal{R}}(0,mx/t) & x>0 \\
 n_{\mathcal{L}} + f_{\mathcal{L}}(0,-mx/t)+ f_{\mathcal{R}}(-mx/t,\infty) & x<0\end{array}\right. ,
 \ee
 where we have introduced the scaling function
 \be
 f_{\mathcal{L/R}}(x,y) \equiv \int_{x}^{y} \frac{dp}{\pi} \tilde n_{\mathcal{L/R}}(p),
 \ee
 and defined the left and right initial particle densities
  \be\label{initial_n}
 n_{\mathcal{L/R}} \equiv f_{\mathcal{L/R}}(0,\infty) = 
 -\left(\frac{m}{2\pi}\right)^{\frac{d}{2}}
 \frac{{\rm Li}_{\frac{d}{2}}\left( -{\rm e}^{\beta_{\mathcal{L/R}}\mu_{\mathcal{L/R}}}\right)  }{\beta_{\mathcal{L/R}}^{\frac{d}{2}}},
 \ee
 where the Polylogarithm function ${\rm Li}_{s}(z)$ is defined in \ref{app_effective_functions}.
 
 Following the same lines, we can describe the time-dependent energy-density profile. In this case, the initial one-dimensional 
 energy distribution function is obtained from Eq. (\ref{E_x_t_1D}) after coarse graining the wave functions over the phase-space.
 Otherwise, the same result can be achieved by considering, from the very beginning, a $2d$-dimensional phase-space and integrating
 out the degrees of freedom orthogonal to $(x_1,p_1)$. Independently on the approach, we finally get 
  \be\label{e0_px}
 \mathcal{E}_{0}(x,p) = \frac{ \tilde\mathcal{E}_{\mathcal{L}}(p) + \tilde n_{\mathcal{L}}(p) p^2/(2m)}{\pi}\theta(-x)\\
 +  \frac{ \tilde\mathcal{E}_{\mathcal{R}}(p) + \tilde n_{\mathcal{R}}(p) p^2/(2m)}{\pi}\theta(x),
 \ee
 from which we straightforwardly obtain
 \be\label{Ext_scaling}
 \mathcal{E}(x,t) = \frac{1}{2}\left\{\begin{array}{cc}  \mathcal{E}_{\mathcal{R}} + g_{\mathcal{L}}(mx/t,\infty)+ g_{\mathcal{R}}(0,mx/t) & x>0 \\
  \mathcal{E}_{\mathcal{L}} + g_{\mathcal{L}}(0,-mx/t)+ g_{\mathcal{R}}(-mx/t,\infty) & x<0\end{array}\right. ,
 \ee
 with 
  \be
 g_{\mathcal{L/R}}(x,y) \equiv \int_{x}^{y} \frac{dp}{\pi} \left[ \tilde\mathcal{E}_{\mathcal{L/R}}(p) 
 + \tilde n_{\mathcal{L/R}}(p) \frac{p^2}{2m}\right],
 \ee
 and initial energy densities
 \be\label{initial_E}
 \mathcal{E}_{\mathcal{L/R}} \equiv g_{\mathcal{L/R}}(0,\infty) = 
   -\frac{d}{2}\left(\frac{m}{2\pi}\right)^{\frac{d}{2}}
 \frac{{\rm Li}_{\frac{d}{2}+1}\left( -{\rm e}^{\beta_{\mathcal{L/R}}\mu_{\mathcal{L/R}}}\right)  }{\beta_{\mathcal{L/R}}^{\frac{d}{2}+1}}. 
 \ee

 In Figures \ref{fig_nE_2D_scaling} and \ref{fig_nE_3D_scaling} 
 we compare the semiclassical predictions (\ref{nxt_scaling})
 and (\ref{Ext_scaling}) with the numerically evaluated particle and energy density
 profiles in two and three dimensions for initial temperatures $\beta_{\mathcal{L}}=1,\;\beta_{\mathcal{R}}=0.5$
  and chemical potentials equals ($\mu_{\mathcal{L}}= \mu_{\mathcal{R}}=1$) or
  different ($\mu_{\mathcal{L}}=2,\; \mu_{\mathcal{R}}=1$). Notice how, in this last situation,
 the particle and energy rescaled profiles present regions with opposite gradient, 
 therefore the particle and energy stationary currents 
 may flow in opposite directions. Moreover, we also consider the particular initial situation
 characterized by the same left and right particle densities 
 (wich corresponds to the rightmost column in Figures \ref{fig_nE_2D_scaling} and \ref{fig_nE_3D_scaling}). 
 Notwithstanding, the stationary state generated by the post-quench dynamics is 
 characterized by a non-vanishing current of particles.

 \section{Particle and energy currents along the quenched direction}\label{sec_currents}
 Using these last results, we can easily derive the particle and the energy currents flowing 
 along the quenched direction and passing the interface separating the two semi-infinite half domains. 
 Indeed, from the continuity equation (\ref{particle_current})
 and using the scaling form in Eq. (\ref{nxt_scaling}), the current of
 particle which flows from $\mathcal{L}$ to $\mathcal{R}$ is given by
 \be
\mathcal{J}(z) = -\int_{-\infty}^{z} \frac{dp}{\pi} \, \frac{p}{2m} \left[ \tilde n_{\mathcal{L}}(p) - \tilde n_{\mathcal{R}}(p)  \right].
 \ee
 in terms of the scaling variable $z=mx/t$, where we used the fact that 
 $\tilde n_{\mathcal{L/R}}(-p)=\tilde n_{\mathcal{L/R}}(p)$. 
 Notice that in the scaling regime, the current at the interface $x=0$ 
  coincides to the non-equilibrium stationary current (for $t\to\infty$) and it does not depend on time as expected. 
  Similarly, injecting Eq. (\ref{Ext_scaling}) in Eq. (\ref{energy_current}), the energy current is obtained
 \be
 \vartheta(z) = -\int_{-\infty}^{z} \frac{dp}{\pi} \, \frac{p}{2m} 
 \left\{  \tilde\mathcal{E}_{\mathcal{L}}(p) -  \tilde\mathcal{ E}_{\mathcal{R}}(p) 
 +  \frac{p^2}{2m} \left[ \tilde n_{\mathcal{L}}(p) - \tilde n_{\mathcal{R}}(p) \right]\right\},
 \ee
 where we used the parity of $\tilde\mathcal{ E}_{\mathcal{L/R}}(p)$.
 
 In particular, making use of the results reported in \ref{app_effective_functions},
 we can explicitly evaluate the particle and energy stationary currents along the quenched direction for 
 any dimension $d$. Indeed, we obtain for the current of particle in the non-equilibrium 
 stationary state $\mathcal{J}_{NESS}\equiv \mathcal{J}(0)$
 \be\label{Jn_NESS}
 \mathcal{J}_{NESS} = \frac{m^{\frac{d-1}{2}}}{(2\pi)^{\frac{d+1}{2}}} \left[ \frac{ -{\rm Li}_{\frac{d+1}{2}} \left( -{\rm e}^{\beta_{\mathcal{L}}\mu_{\mathcal{L}}}\right) }{\beta_{\mathcal{L}}^{\frac{d+1}{2}}} 
 - \frac{ -{\rm Li}_{\frac{d+1}{2}} \left( -{\rm e}^{\beta_{\mathcal{R}}\mu_{\mathcal{R}}}\right) }{\beta_{\mathcal{R}}^{\frac{d+1}{2}}}\right].
 \ee
 In the same way, the stationary energy current $\vartheta_{NESS} \equiv \vartheta(0)$ is given by
  \be\label{JE_NESS}
 \vartheta_{NESS} = \frac{d+1}{2}\frac{m^{\frac{d-1}{2}}}{(2\pi)^{\frac{d+1}{2}}} \left[ \frac{ -{\rm Li}_{\frac{d+3}{2}} \left( -{\rm e}^{\beta_{\mathcal{L}}\mu_{\mathcal{L}}}\right) }{\beta_{\mathcal{L}}^{\frac{d+3}{2}}} 
 - \frac{ -{\rm Li}_{\frac{d+3}{2}} \left( -{\rm e}^{\beta_{\mathcal{R}}\mu_{\mathcal{R}}}\right) }{\beta_{\mathcal{R}}^{\frac{d+3}{2}}}\right].
 \ee
These latter equations can be expanded at low temperatures. Indeed,
using the asymptotic expansion of the Polylogarithm functions,
one obtains, for $\mu_{\mathcal{L/R}} > 0$, up to $O(1/\beta_{\mathcal{L/R}}^{2})$
\be\label{Jn_lowT}
\mathcal{J}_{NESS} =
\frac{m^{\frac{d-1}{2}}}{(2\pi)^{\frac{d+1}{2}}} \left[
\frac{\mu_{\mathcal{L}}^{\frac{d+1}{2}}-\mu_{\mathcal{R}}^{\frac{d+1}{2}}}{\Gamma(\frac{d+3}{2})}
+\frac{\pi^{2}}{6\,\Gamma(\frac{d-1}{2})}
\left(\frac{\mu_{\mathcal{L}}^{\frac{d-3}{2}}}{\beta_{\mathcal{L}}^{2}}-\frac{\mu_{\mathcal{R}}^{\frac{d-3}{2}}}{\beta_{\mathcal{R}}^{2}}\right) \right],
\ee
and
\be\label{JE_lowT}
 \vartheta_{NESS} = 
 \frac{(d+1)m^{\frac{d-1}{2}}}{2(2\pi)^{\frac{d+1}{2}}}  \left[
 \frac{\mu_{\mathcal{L}}^{\frac{d+3}{2}}-\mu_{\mathcal{R}}^{\frac{d+3}{2}}}{\Gamma(\frac{d+5}{2})}
 +\frac{\pi^{2}}{6\,\Gamma(\frac{d+1}{2})}
 \left(\frac{\mu_{\mathcal{L}}^{\frac{d-1}{2}}}{\beta_{\mathcal{L}}^{2}}-\frac{\mu_{\mathcal{R}}^{\frac{d-1}{2}}}{\beta_{\mathcal{R}}^{2}}\right) \right],
\ee
where $\Gamma(z)$ is the usual Gamma function. Notice that, since $\Gamma(0)$ is diverging,
the low-temperature behavior of the stationary particle current $\mathcal{J}_{NESS}$ at $d=1$
depends only on the chemical potential gradient, in agreement with the asymptotic expansion 
of ${\rm Li_{1}(z)}$.

Nevertheless, for all other dimensions, at low temperatures, and at fixed equal chemical potentials, 
both the particle and the energy currents show the same quadratic behavior in temperature. 

\subsection{Non-equilibrium transport for uniform initial particle densities}
Often, in the experimental or numerical setups, the particle and energy transport characterizing the
non-equilibrium stationary state arises from fixed initial densities of particle. In other words, it would be
interesting to analyze the scaling properties of the previous currents whenever the initial conditions
of the left and right regions are such that, the initial densities of particle are fixed and equals. 
 Since the temperatures $1/\beta_{\mathcal{L/R}}$ are still free parameters, the condition on the 
 initial densities becomes a condition on the chemical potentials $\mu_{\mathcal{L/R}}$. In particular,
 fixing the initial left and right densities of particles  to $n_{0}$, and  expanding Eq. (\ref{initial_n}) 
 in the low temperature regime ($\beta\gg 1$), one has
 \be
 n_0 \left( \frac{2\pi}{m}\right)^{d/2} = \frac{\mu^{d/2}}{\Gamma(d/2+1)}
 + \frac{\pi^2}{6\,\Gamma(d/2-1)}\frac{\mu^{d/2-2}}{\beta^2}
  + O\left(\frac{1}{\beta^4}\right),
 \ee
 which can be iteratively solved for $\mu(\beta)$ in power of $1/\beta$, giving
 \be\label{mu_expans}
 \mu(\beta) = \mu_{0}\left[1-\frac{(d-2)\pi^2}{12\,\mu_0^2}\frac{1}{\beta^2}\right]  + O\left(\frac{1}{\beta^4}\right),
 \ee
 with $\mu_0 = 2\pi [n_0 \Gamma(d/2+1)]^{2/d}/m$. Notice that, for $d=2$ the second order 
 correction on the chemical potential expansion is zero and the first non-vanishing correction will 
 be of order $1/\beta^4$. However, thanks to the structure of the equations  (\ref{Jn_lowT})
 and  (\ref{JE_lowT}), for any dimensions $d$ there will always be a non-zero contribution 
 dependent on the square of the temperatures. 
 Indeed, substituting the latter expansion in the Eq. (\ref{Jn_lowT})
 and Eq. (\ref{JE_lowT}), and retaining the terms up to $O(1/\beta_{\mathcal{L/R}}^2)$, one finally gets
 \bea
 \mathcal{J}|_{n_0} & = & \frac{\pi^2}{12\,\Gamma(\frac{d+1}{2})} \frac{m^{\frac{d-1}{2}} }{(2\pi)^{\frac{d+1}{2}}} 
  \,\mu_0^{\frac{d-3}{2}} \left( \frac{1}{\beta_{\mathcal{L}}^2} -  \frac{1}{\beta_{\mathcal{R}}^2}  \right) \\
  & = & \frac{m}{48}\frac{\Gamma(\frac{d}{2}+1)^{\frac{d-3}{d}}}{\Gamma(\frac{d+1}{2})}
   \, n_0^{\frac{d-3}{d}}  \left( \frac{1}{\beta_{\mathcal{L}}^2} -  \frac{1}{\beta_{\mathcal{R}}^2}  \right)
\eea
and
\bea
 \vartheta|_{n_0} & = & \frac{\pi^2}{4\,\Gamma(\frac{d+1}{2})} \frac{m^{\frac{d-1}{2}} }{(2\pi)^{\frac{d+1}{2}}}
 \, \mu_0^{\frac{d-1}{2}}  \left( \frac{1}{\beta_{\mathcal{L}}^2} -  \frac{1}{\beta_{\mathcal{R}}^2}  \right) \\
  & = & \frac{1}{8\pi}\frac{\Gamma(\frac{d}{2}+1)^{\frac{d-1}{d}}}{\Gamma(\frac{d+1}{2})}
   \, n_0^{\frac{d-1}{d}}  \left( \frac{1}{\beta_{\mathcal{L}}^2} -  \frac{1}{\beta_{\mathcal{R}}^2}  \right).
 \eea
 As expected in this particular regime, the zero order contribution vanishes, 
 and both the currents show an universal behavior proportional to the 
 difference $T_{\mathcal{L}}^{2}-T_{\mathcal{R}}^{2}$,
 similarly to the $3D$ calculation in Ref. \cite{gk42}.
Let us mention also that, having prepared the system with an uniform initial particle density did not prevent 
the generation of a non-vanishing current of particles in the non-equilibrium stationary-state.
This is a direct consequence of what we have already shown 
for the stationary particle-density profile in Figures \ref{fig_nE_2D_scaling} 
and \ref{fig_nE_3D_scaling}.

\section{The massless relativistic case}\label{RC}
We now consider massless relativistic free fermions with dispersion relation
\be
\epsilon({\bm k})=v_{F}|{\bm k}|,
\ee
with ${\bm k}=\{k_1,\dots k_d\}$.
We study the relativistic case in order to emphasize that the low 
temperature expansion of the currents for the non-relativistic case does not 
coincide with the analogous problem for relativistic massless fermions. 
Indeed, even though for small temperatures and near the Fermi surface 
the non-relativistic dispersion relation can be linearized, only in $1D$ 
one recovers exactly the same scaling behavior of the massless relativistic case.  

Using the same approach of the previous sections, we obtain the stationary particle current
along the quenched direction $ x_1$
\bea
\mathcal{J}_{rel} & = & -\frac{1}{2}\int^{0}_{-\infty} \frac{d^{d}k}{\pi^d} \frac{d\epsilon}{d k_1}
\left[ n_{\mathcal{L}}({\bm k}) - n_{\mathcal{R}}({\bm k})  \right] \\
& = & -\frac{1}{2}\int_{-\infty}^{0} \frac{d^{d}k}{\pi^{d}} \, v_{F}\cos(\theta) 
\left[ n_{\mathcal{L}}({\bm k}) - n_{\mathcal{R}}({\bm k})  \right],
 \eea
 where $n({\bm k})=[\exp(\beta v_{F}|{\bm k}|)+1]^{-1}$ and 
 we used the fact that $d\epsilon/d k_{1}=v_F k_{1}/|{\bm k}| = v_{F}\cos(\theta)$.
Similarly, the energy current is obtained
\begin{small}
\bea
\vartheta_{rel} & =&  -\frac{1}{2}\int_{-\infty}^{0} \frac{d^{d}k}{\pi^d} \,  \frac{d\epsilon}{d k_1}
\epsilon({\bm k}) \left[ n_{\mathcal{L}}({\bm k}) - n_{\mathcal{R}}({\bm k}) \right]\\
& = & -\frac{1}{2}\int_{-\infty}^{0} \frac{d^{d}k}{\pi^d} \, v^{2}_{F}\cos(\theta)
|{\bm k}| \left[ n_{\mathcal{L}}({\bm k}) - n_{\mathcal{R}}({\bm k}) \right].
\eea
\end{small}
Explicitly evaluating the previous integrals, one has for the
non-equilibrium stationary current of particles
\be\label{Jn_rel}
\mathcal{J}_{rel}=\frac{(2^{d-1}-1)\Gamma(d)\zeta(d)}{(2\pi)^{d}\Gamma(\frac{d+1}{2})v_{F}^{d-1}} \left(\frac{1}{\beta_{\mathcal{L}}^{d}}-\frac{1}{\beta_{\mathcal{R}}^{d}}\right),
\ee
where $\zeta(z)$ is the Riemann zeta function. In the same way, we obtain for the energy current
\be\label{JE_rel}
\vartheta_{rel}=\frac{(2^{d}-1)\Gamma(d+1)\zeta(d+1)}{2(2\pi)^{d}\Gamma(\frac{d+1}{2})v_{F}^{d-1}}\left(\frac{1}{\beta_{\mathcal{L}}^{d+1}}-\frac{1}{\beta_{\mathcal{R}}^{d+1}}\right).
\ee
As expected, the relativistic currents, for $d=1$, perfectly match the non-relativistic results in
(\ref{Jn_NESS}) and (\ref{JE_NESS}) evaluated at zero chemical potentials, i.e.
\be
\mathcal{J}_{rel}|_{d=1} = \frac{\log(2)}{2\pi} 
\left(\frac{1}{\beta_{\mathcal{L}}}-\frac{1}{\beta_{\mathcal{R}}}\right),\quad
\vartheta_{rel}|_{d=1} = \frac{\pi}{24} 
\left(\frac{1}{\beta^{2}_{\mathcal{L}}}-\frac{1}{\beta^{2}_{\mathcal{R}}}\right),
\ee 
which agree with what was already found in Ref. \cite{ck14}.

Nonetheless, some comments about Eq.s (\ref{Jn_rel}) and (\ref{JE_rel}) are due. 
Comparing these expressions with equation (8) in Ref. \cite{HolograficDoyon}, 
some differences are evident. Although the explicit temperature behaviors
are exactly the same, a different dependence on the left and right velocities appears. 
Such a difference would still remain even if one considers $v_{F}$ being different 
in the two halves and depending on temperature.

We argue that this difference is present since 
we are considering opposite regimes: in Ref. \cite{HolograficDoyon} the authors 
take into account an ideal relativistic fluid strongly coupled in the hydrodynamic limit,
whereas we studied a free theory which has, by definition, an infinite mean free path.

\section{Conlusions}\label{Concl}
In this paper we studied the particles and energy currents 
in the non-equilibrium stationary-state of a $d$-dimensional Fermi gas 
initially prepared into two halves at different temperatures and chemical potentials. 
After having joined the two halves with a local coupling, we left the system 
to evolve with a non-interacting Hamiltonian. 
We exactly characterized the dynamics of the
particles and energy profiles by means of a semiclassical 
approach \cite{akr08,cark12,wck13}, 
and we found a perfect matching with the exact numerical computation.
Moreover, for generic spatial dimensions $d$, we analytically computed 
the steady-state particle and energy currents. 

In particular, for a non-relativistic fermion gas, 
the exact expression for the particle and energy currents strongly depends
to the dimensionality $d$ of the system. Nevertheless, we observed an 
universal transport behavior at low temperatures proportional
to $(T_{\mathcal{L}}^{2}-T_{\mathcal{R}}^{2})$. 

Moreover, we analyzed the difference between the non-relativistic and the massless 
relativistic case, and we observed that
 the behavior of the non-relativistic case near the Fermi surface, i.e. for $T_{\mathcal{L/R}}\ll 1$, 
 is different to the relativistic calculation.
 In this regard, we stressed that only for $d=1$ the two cases coincide.
 
 In other words, the non-equilibrium transport properties of a non-relativistic 
 Fermi gas in $d>1$ do not show a conformal invariant behavior.

\section{Acknowledgements}
The authors thank F. Bigazzi, P. Calabrese, M. Mintchev and E. Vicari for fruitful discussions
and acknowledge the ERC for financial support under Starting Grant 279391 EDEQS.

\appendix
\section{Effective mode occupation and energy distribution function}\label{app_effective_functions}
The functions characterizing the one-dimensional equivalent model, i. e. $\tilde n_{\mathcal{L/R}}(p)$
and $\tilde\mathcal{E}_{\mathcal{L/R}}(p)$ can be explicitly evaluated for any original dimension $d$.
Indeed, using the $(d-1)$-dimensional spherical coordinates, the mode occupation can be written as
\be
\tilde n_{\mathcal{L/R}}(p) = \frac{S_{d-2}}{(2\pi)^{d-1}}\int_{0}^{\infty} 
\frac{\rho^{d-2} d\rho}{1+{\rm e}^{\beta_{\mathcal{L/R}}[(p^2+\rho^2) /(2m)-\mu_{\mathcal{L/R}}]}},
\ee
where $S_{d-2} = (d-1) \pi^{(d-1)/2}/\Gamma[(d+1)/2]$ is the surface of the $(d-1)$-dimensional sphere with
unitary radius. In particular, the radial integral can be explicitly evaluated in terms of Polylogarithm functions
${\rm Li}_{s}(z)\equiv \sum_{k=1}^{\infty}z^k / k^s$, and finally one obtains
\be
\tilde n_{\mathcal{L/R}}(p) = -\left( \frac{m}{2\pi}\right)^{\frac{d-1}{2}} 
\frac{ {\rm Li}_{\frac{d-1}{2}}\left( -{\rm e}^{\beta_{\mathcal{L/R}}[\mu_{\mathcal{L/R}}-p^2/(2m)]}  \right)}{\beta_{\mathcal{L/R}}^{\frac{d-1}{2}}}.
\ee
Similarly, the energy distribution function is given by
\be
\tilde\mathcal{E}_{\mathcal{L/R}} (p) = -\frac{d-1}{2}\left( \frac{m}{2\pi}\right)^{\frac{d-1}{2}} 
\frac{ {\rm Li}_{\frac{d+1}{2}}\left( -{\rm e}^{\beta_{\mathcal{L/R}}[\mu_{\mathcal{L/R}}-p^2/(2m)]}  \right)}{\beta_{\mathcal{L/R}}^{\frac{d+1}{2}}}.
\ee
We conclude this appendix giving some useful formulae regarding the Polylogarithm functions.
The following integral will be useful in the main text in order to evaluate the stationary currents:
 \bea
  && \int_{-\infty}^{0}dp \, p^\alpha \, {\rm Li}_{s}\left( -{\rm e}^{\beta\mu}{\rm e}^{-\beta p^2 /(2m)}\right) = \\
 &&= \frac{(-1)^{\alpha}\Gamma[(\alpha+1)/2]}{2}\left( \frac{2m}{\beta} \right)^{\frac{\alpha+1}{2}} {\rm Li}_{s + \frac{\alpha+1}{2}}\left(-{\rm e}^{\beta\mu}\right).\nonumber
 \eea
 Furthermore, in order to extract the small temperature behavior, we will make often use of the following
 asymptotic expansion for $|z|\gg 1$ valid for all $s$ and any $\arg(z)$
 \be
 {\rm Li}_{s}\left(-{\rm e}^{z}\right) = \sum_{k=0}^{\infty} \frac{(-1)^k (1-2^{1-2k})(2\pi)^{2k} B_{2k} }{(2k)! \Gamma[s+1-2k]} z^{s-2k},
 \ee
 where $B_{2k}$ are the Bernoulli numbers.




\Bibliography{99}
\addcontentsline{toc}{section}{References}

\bibitem{uc}
M. Greiner, O. Mandel, T. W. H\"ansch, and I. Bloch, Nature {\bf 419} 51 (2002).

\bibitem{kww06}
T. Kinoshita, T. Wenger,  D. S. Weiss, Nature {\bf 440}, 900 (2006).

\bibitem{tc07}
S. Hofferberth, I. Lesanovsky, B. Fischer, T. Schumm, and J. Schmiedmayer, Nature {\bf 449}, 324 (2007).

\bibitem{tetal11}
S. Trotzky Y.-A. Chen, A. Flesch, I. P. McCulloch, U. Schollw\"ock, J. Eisert, and I. Bloch, Nature Phys. {\bf 8}, 325 (2012). 

\bibitem{cetal12}
M. Cheneau, P. Barmettler, D. Poletti, M. Endres, P. Schauss, T. Fukuhara, C. Gross, I. Bloch, C. Kollath, and S. Kuhr, Nature {\bf 481}, 484 (2012).

\bibitem{getal11}
M. Gring, M. Kuhnert, T. Langen, T. Kitagawa, B. Rauer, M. Schreitl, I. Mazets, D. A. Smith, E. Demler, and J. Schmiedmayer, Science {\bf 337}, 1318 (2012).

\bibitem{shr12}
U. Schneider, L. Hackerm\"uller, J. P. Ronzheimer, S. Will, S. Braun, T. Best, I. Bloch, E. Demler, S. Mandt, D. Rasch, and A. Rosch, Nature Phys. {\bf 8}, 213 (2012).

\bibitem{rsb13}
J. P. Ronzheimer, M. Schreiber, S. Braun, S. S. Hodgman, S. Langer, I. P. McCulloch, F. Heidrich-Meisner, I. Bloch, and U. Schneider, Phys. Rev. Lett. {\bf 110}, 205301 (2013).

\bibitem{rdyo07} 
M. Rigol, V. Dunjko, V. Yurovsky,  and M. Olshanii, Phys. Rev. Lett. {\bf 98}, 50405 (2007);
M. Rigol, V. Dunjko,  and M. Olshanii, Nature {\bf 452}, 854 (2008). 


\bibitem{mussardo10-13}
D. Fioretto and G. Mussardo New J. Phys. {\bf 12} 055015 (2010);
S. Sotiriadis, D. Fioretto, G. Mussardo, J. Stat. Mech. P02017 (2012);
G. Mussardo, Phys. Rev. Lett. {\bf 111}, 100401 (2013);

\bibitem{cc06} 
P. Calabrese and  J. Cardy, Phys. Rev. Lett. {\bf 96}, 136801 (2006); 
J. Stat. Mech. P06008  (2007); 
J. Stat. Mech. P04010 (2005).

\bibitem{c06}
M. A. Cazalilla, Phys. Rev. Lett. {\bf 97}, 156403 (2006); 
A. Iucci, and M. A. Cazalilla, Phys. Rev. A {\bf 80}, 063619 (2009); New J. Phys. {\bf 12}, 055019 (2010);
A. Mitra and T. Giamarchi, Phys. Rev. Lett. {\bf 107}, 150602 (2011).

\bibitem{cdeo08}
M. Cramer, C. M. Dawson, J. Eisert, and T. J. Osborne, Phys. Rev. Lett. {\bf 100}, 030602 (2008);
M. Cramer and J. Eisert, New J. Phys. 12, 055020 (2010).

\bibitem{bs08}
T. Barthel and U. Schollw\"ock, Phys. Rev. Lett. {\bf 100}, 100601 (2008).

\bibitem{ce13}
J.-S. Caux, F. Essler, Phys. Rev. Lett. {\bf 110} 257203 (2013).

\bibitem{fe13}
M.  Fagotti and  F. H. L. Essler, Phys. Rev. B {\bf 87}, 245107 (2013).

\bibitem{fcec13}
M. Fagotti, M. Collura, F. H. L. Essler and P. Calabrese, Phys. Rev. B {\bf 89}, 125101 (2014).


\bibitem{revq}
A. Polkovnikov, K. Sengupta, A. Silva, and M. Vengalattore, Rev. Mod. Phys. {\bf 83}, 863 (2011).


\bibitem{araki}
H. Araki and T. G. Ho, Proc. Steklov Inst. Math. {\bf 228}, 191 (2000).

\bibitem{ogata}
Y. Ogata, Phys. Rev E {\bf 66}, 066123 (2002); Phys. Rev. E {\bf 66}, 016135 (2002).

\bibitem{aschbacher}
W. H. Aschbacher and C.-A. Pillet, J. Stat. Phys. {\bf 112}, 1153 (2003).

\bibitem{pk07}
D. Karevski, Eur. Phys. J. B {\bf  27}, 147 (2002);
T. Platini, D. Karevski, Eur. Phys. J. B {\bf 48} 225 (2005);
T. Platini, D. Karevski,  J. Phys. A: Math. Theor. {\bf 40}, 1711 (2007).

\bibitem{dvbd13}
A. De Luca, J. Viti, D. Bernard and B Doyon, Phys. Rev. B {\bf 88}, 134301 (2013).

\bibitem{mm11}
M. Mintchev, J. Phys. A {\bf 44}, 415201 (2011).

\bibitem{arrs98} 
T. Antal, Z. R\'acz, A. R\'akos, and G.M. Sch\"utz, Phys. Rev. E {\bf 57}, 5184 (1998).

\bibitem{ruelle98}
 D. Ruelle, J. Stat. Phys. {\bf 98}, 57 (2000).

\bibitem{jp02} 
V. Jak\v{s}i\'cand and C.-A. Pillet, Commun. Math. Phys. {\bf 226}, 131 (2002).

\bibitem{kp09}
D. Karevski, T. Platini, Phys. Rev. Lett. {\bf 102}, 207207 (2009).

\bibitem{prosen11}
T. Prosen, Phys. Rev. Lett. {\bf 107}, 137201 (2011).

\bibitem{kps13}
D. Karevski, V. Popkov, G. M. Schuetz, Phys. Rev. Lett. {\bf 110} 047201(2013);
V. Popkov, D. Karevski, G. M. Schuetz, Phys. Rev. E {\bf 88} 062118 (2013).


\bibitem{cacdh13}
O. Castro-Alvaredo, Y. Chen, B. Doyon, M. Hoogeveen,
J. Stat. Mech. P03011 (2014).
 
 \bibitem{CICC}
 M. A. Cazalilla, A. Iucci, and M.-C. Chung, Phys. Rev. E {\bf 85}, 011133 (2012).

 \bibitem{KIM}
 C. Karrasch, R. Ilan, and J. E. Moore, Phys. Rev. B {\bf 88}, 195129 (2013).
 
 \bibitem{SM}
 T. Sabetta and G. Misguich, Phys. Rev. B {\bf 88}, 245114 (2013).
 
\bibitem{bd12}
D. Bernard and B. Doyon, J. Phys. A {\bf 45}, 362001 (2012); arXiv:1302.3125.
 

\bibitem{kbm12}
C. Karrasch, J. H. Bardarson, and J. E. Moore, 
Phys. Rev. Lett. {\bf 108}, 227206 (2012).

\bibitem{kbm13}
C. Karrasch, J. H. Bardarson, and J. E. Moore, 
New J. Phys. {\bf 15}, 083031 (2013).

\bibitem{hkm13}
Y. Huang, C. Karrasch, and J. E. Moore, 
Phys. Rev. B {\bf 88}, 115126 (2013).

\bibitem{Zamo90}
A. Zamolodchikov, Nucl. Phys. B {\bf 342}, 695 (1990).
 
 
 \bibitem{Hartnoll}
S. A. Hartnoll, Class. Quant. Grav. {\bf 26}, 224002 (2009).

 \bibitem{MG}
J. McGreevy, 
Advances in High Energy Physics, 723105 (2010).

\bibitem{HolograficDoyon}
M. J. Bhaseen, B. Doyon, A. Lucas, and K. Schalm, arXiv:1311.3655.

\bibitem{ck14}
M. Collura and D. Karevski, arXiv:1402.1944.


\bibitem{akr08}
T. Antal, P. L. Krapivsky, and A. R\'akos, 
Phys. Rev. E {\bf 78}, 061115 (2008).

\bibitem{cark12}
M. Collura, H. Aufderheide, G. Roux and D. Karevski, Phys. Rev. A {\bf 86}, 013615 (2012).

\bibitem{wck13}
P. Wendenbaum, M. Collura and D. Karevski, Phys. Rev. A {\bf 87}, 023624 (2013).


\bibitem{localCFT}
P. Calabrese, J. Cardy, J. Stat. Mech. P10004 (2007). 

\bibitem{localentropy}
V. Eisler, and I. Peschel, J. Stat. Mech. P06005  (2007);
V. Eisler, D. Karevski, T. Platini, I. Peschel, J. Stat. Mech. (2008) P01023

\bibitem{peschel05}
I. Peschel, J. Phys. A {\bf 38}, 4327 (2005);
\bibitem{entr_cc13}
M. Collura, P. Calabrese, J. Phys. A {\bf 46}, 175001 (2013).

\bibitem{mpc10}
J. Mossel, G. Palacios, J.-S. Caux, J. Stat. Mech. L09001 (2010). 

\bibitem{ah14}
V. Alba, F. Heidrich-Meisner, arXiv:1402.2299.

\bibitem{gk42}
D. V. Gogate, and D. S. Kothari, Phys. Rev. {\bf 61}, 349 (1942).





\end{thebibliography}
\end{document}